\documentclass{aa}    

\usepackage{upgreek}

\usepackage[varg]{txfonts}
\usepackage{hyperref}
\usepackage{graphicx}

\hypersetup{%
  colorlinks=true,
  linkcolor=blue,
  citecolor=blue
}

\usepackage{etoolbox}

\let\oldcitet=\citet

\renewcommand{\citet}[1]{\textcolor[rgb]{0,0,1}{\oldcitet{#1}}}

\begin{document}

\title{Anomalous HCN emission from warm giant molecular clouds}

 \titlerunning{Anomalous HCN emission from warm GMCs} 
\authorrunning{Goicoechea et al.} 
                                      
 \author{Javier R.\,Goicoechea\inst{1}
          \and
        Fran\c{c}ois Lique\inst{2}
        \and
        Miriam G. Santa-Maria\inst{1}}

\institute{Instituto de F\'{\i}sica Fundamental
     (CSIC). Calle Serrano 121-123, 28006, Madrid, Spain. 
     \email{javier.r.goicoechea@csic.es}
\and
  Univ. Rennes, CNRS, IPR (Institut de Physique de Rennes), UMR 6251, F-35000 Rennes, France.    
}

   \date{Received 13 September 2021 / Accepted 30 October 2021}



\abstract{Hydrogen cyanide (HCN)  is  considered a good tracer of the dense molecular gas that serves as fuel for star formation. However, recent large-scale surveys of   giant molecular clouds (GMCs) have  detected extended HCN rotational line emission far from  star-forming cores.  
Such observations often spectroscopically  resolve the  \mbox{HCN $J$\,=\,1--0} 
(partially also the $J$\,=\,2--1 and 3--2)
\mbox{hyperfine} structure (HFS). 
A precise determination of the physical conditions of the gas requires treating the HFS line overlap  effects.
 Here, we study the HCN HFS excitation and line emission 
 using nonlocal radiative transfer models that include line overlaps and
 new HFS-resolved \mbox{collisional} rate coefficients  for inelastic collisions of HCN
with both \mbox{para-H$_2$} and \mbox{ortho-H$_2$} (computed via the 
\mbox{scaled-infinite order sudden  approximation}    up to \mbox{$T_{\rm k}$\,=\,500\,K}).
In addition, we account for the role of electron collisions in the HFS level excitation.
We find that  
line overlap and opacity effects frequently  produce anomalous \mbox{HCN $J$\,=\,1--0} HFS line intensity ratios \mbox{(i.e., inconsistent with the common assumption of the  same $T_{\rm ex}$ for all HFS lines)}
as well as anomalous HFS line width ratios. 
Line overlap and electron collisions also enhance the excitation of  the
higher $J$ rotational lines. Our models  explain the anomalous \mbox{HCN $J$\,=\,1--0} HFS spectra observed 
in the  Orion Bar and Horsehead photodissociation regions.
 As shown in previous studies, electron excitation becomes important for
 molecular gas with H$_2$ densities below a few 10$^5$\,cm$^{-3}$ and electron abundances above \mbox{$\sim$\,10$^{-5}$}. We find that when electron collisions are dominant, the relative intensities
of the \mbox{HCN $J$\,=\,1--0} HFS lines can be anomalous too.
In particular, electron excitation can produce low-surface-brightness  \mbox{HCN} emission from  very extended but low-density  gas in  GMCs.
The existence of such a widespread HCN emission component   may affect
the interpretation of the  extragalactic relationship HCN luminosity versus star-formation rate.
 Alternatively,  extended HCN emission may arise from dense star-forming cores
 and become resonantly scattered
by large envelopes of lower density gas. 
There are two scenarios -- namely, electron-assisted 
(weakly) collisionally  excited versus
scattering -- that lead to different \mbox{HCN $J$\,=\,1--0} HFS
intensity ratios, which can be  tested on the basis of observations. 
}


 \maketitle
%

\section{Introduction}

Massive star clusters form within dense clumps inside giant molecular clouds \citep[GMCs; ][]{Lada03}. Finding appropriate spectroscopic tracers of the
dense molecular gas (\mbox{$n_{\rm H}$= $n$(H)\,+\,2$n$(H$_2$)\,$\gtrsim$\,10$^5$\,cm$^{-3}$}  and linking their line luminosity with the star formation rate is of critical importance  \mbox{\citep[][]{Gao04,Elmegreen07}}. Rotational  emission from hydrogen cyanide (HCN) has long been considered as an appropriate probe for such gas reservoirs feeding
for star formation. Indeed, 
HCN is  sufficiently abundant  to be detected in nearby star-forming clumps as well as in distant galaxies.  
With a dipole moment ($\mu_{\rm e}$) $\sim$30 times higher than that of CO 
(0.11\,D versus 2.99\,D), collisional excitation
of HCN rotational levels requires higher H$_2$ densities than those needed to excite CO.

Paradoxically, recent  surveys of local GMCs, mapping star-forming clumps as well as  their extended cloud 
environment, have revealed widespread  \mbox{HCN $J$\,=\,1--0} emission \citep[e.g.,][]{Pety17,Kauffmann17,Nishimura17, Shimajiri17,Evans20}
associated
with relatively low \mbox{visual} extinctions ($A_V$); thus, it is shown to be associated with gas that may not be so dense, at several 10$^3$\,cm$^{-3}$.
  Indeed,  \mbox{translucent} 
clouds \mbox{($A_V$\,$\lesssim$\,5\,mag)} show HCN emission consistent with HCN abundances up to \mbox{$\chi$(HCN)\,$\simeq$\,10$^{-8}$} \mbox{\citep{Turner97}}. Even the lowest density \mbox{diffuse} molecular clouds (\mbox{$A_V$\,$\lesssim$\,1\,mag}) show \mbox{HCN $J$\,=\,1--0} absorption lines \mbox{\citep[e.g.,][]{Liszt01,Godard10}}
that are compatible with the HCN abundances inferred  in dense clouds.

  Very polar \mbox{neutral} molecules  have large cross sections for inelastic collisions with
electrons, that roughly scale according to $\mu_{\rm e}^{2}$ \citep[][]{Faure07}. This leads to 
collisional rate coefficients that are at least three to four orders of magnitude 
 greater that those induced by collisions with neutral species.
 Electron collisions contribute to 
the excitation of interstellar 
HCN when the electron abundance
(the so-called \mbox{ionization fraction}, \mbox{$\chi_{\rm e}$\,=\,$n_{\rm e}$/$n_{\rm H}$}) 
 is $\geq$\,10$^{-5}$ and the H$_2$ density is not especially high 
 \citep[\mbox{$<$ a few 10$^{5}$\,cm$^{-3}$}; e.g.,][]{Dickinson77,Goldsmith17}. For
 instance, electron collisions control 
 the weak HCN rotational excitation in diffuse  clouds \citep[][]{Liszt12}.  

Owing to the clumpy or fractal structure of dense molecular clouds \mbox{\citep[e.g.,][]{Stutzki90,Falgarone91}}, the  extended component of GMCs is porous to ultraviolet (UV) \mbox{radiation} from nearby massive stars \citep[e.g.,][]{Boisse90}. 
The maximum ionization fraction  in ordinary GMCs \mbox{appears} in the first \mbox{$\approx$\,2-3\,mag} of visual extinction into the  neutral cloud \mbox{\citep[e.g.,][]{Hollenbach91}}.
At these low values for $A_V$, most  electrons  arise from the photoionization of carbon atoms; hence,
 $\chi_{\rm e}$\,$\simeq$\,$\chi$(C$^+$);
with \mbox{$\chi$(C$^+$)\,$\simeq$\,1.4$\cdot$10$^{-4}$} in Orion \citep{Sofia_2004}.
The rims of dense molecular clouds (their photodissociation regions or PDRs) and the  spatially extended GMC environment, where $A_V$ and $n_{\rm H}$
naturally drop, exhibit high ionization fractions.
 At intermediate cloud depths, from \mbox{$A_V$\,$\approx$\,2-3\,mag} to
\mbox{$\approx$\,4-5\,mag,} depending on the gas density and UV photon flux, the ionization fraction is
controlled by the gas-phase abundance of lower ionization potential elements, sulfur in particular; thus, \mbox{$\chi_{\rm e}$\,$\simeq$\,$\chi$(S$^+$)}.  Observations of \mbox{sulfur} radio recombination lines  imply \mbox{$\chi$(S$^+$)\,$\simeq$\,1.4$\cdot$10$^{-5}$} in the Orion Bar PDR \mbox{\citep{Goicoechea21}}.
 Deeper inside the  cloud, in the cold cores where star-formation actually takes place, the ionization fraction  
decreases to  \mbox{$\chi_{\rm e}$\,$\simeq$\,10$^{-8}$}, 
and $n_{\rm e}$ is negligibly low  \citep[e.g.,][]{Guelin82,Caselli98,Maret07, Goicoechea09,Bron21}.

\begin{figure}[t]
\centering   
\includegraphics[scale=0.41, angle=0]{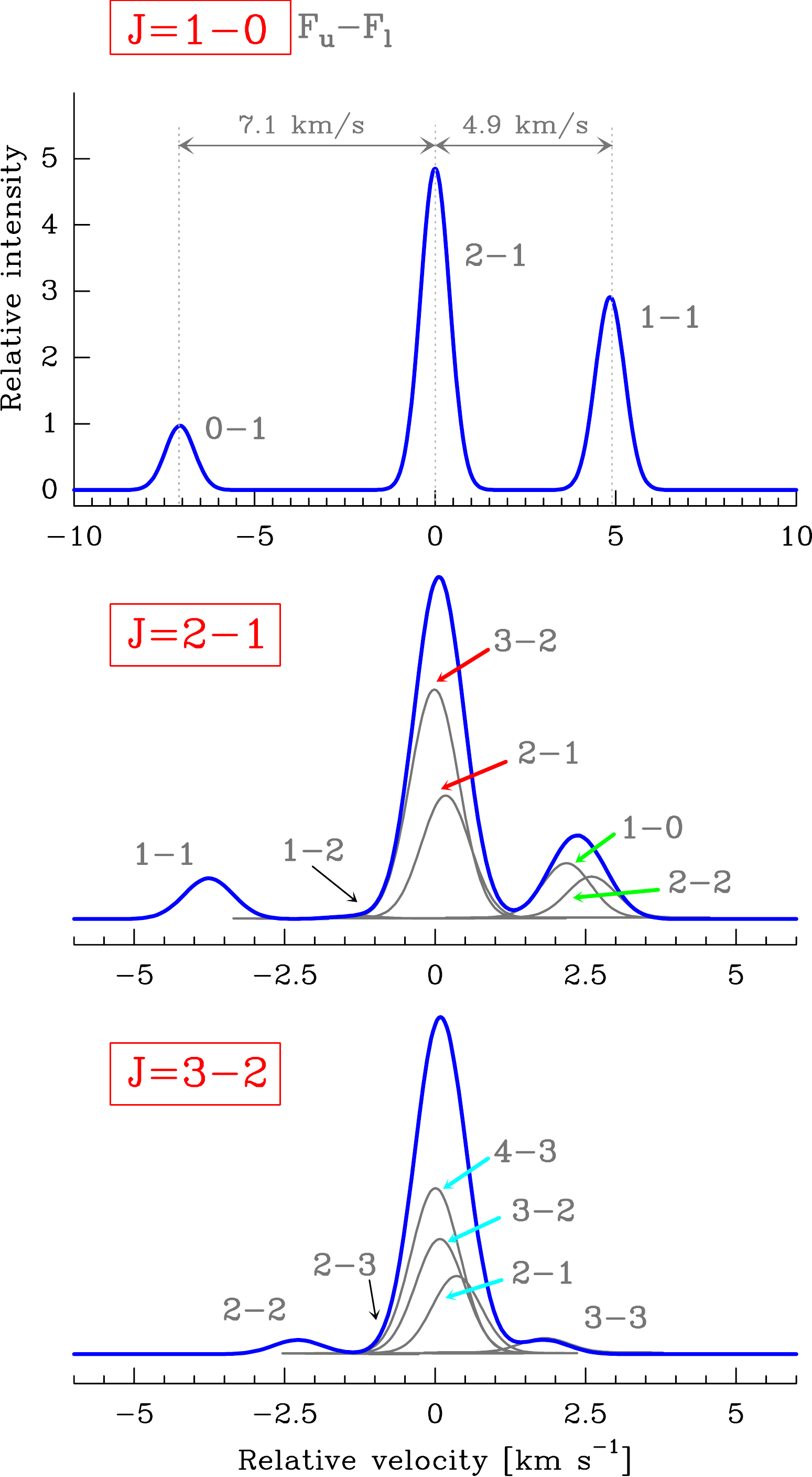}
\caption{Optically thin HCN \mbox{$J$\,=\,1--0} ($\sim$88.6\,GHz), 
\mbox{2--1} ($\sim$177.3\,GHz)
and  \mbox{3--2} ($\sim$265.9\,GHz) spectra for a cloud with \mbox{$\Delta$$v$\,$\simeq$\,1\,km\,s$^{-1}$}.
Each panel shows the velocity separation from the strongest HFS component.
As line opacities increase, line overlaps
in the \mbox{$J$\,=\,2--1} and \mbox{3--2} HFS  transitions affect
the global excitation of HFS levels. 
Red, green, and cyan arrows show specific lines that overlap and produce most of the anomalous HFS emission discussed in the text (see also Fig.~\ref{fig:energy_diagram}).} 
\label{fig:lines_sketch}
\end{figure}

Because of the nitrogen atom, HCN rotational levels possess hyperfine structure (HFS). The coupling between the nuclear spin ($I=1$ for $^{14}$N)  and the molecular rotation 
splits each rotational level $J$ into three hyperfine levels (except  level \mbox{$J$\,=\,0}). Each hyperfine level  is designated by a quantum number 
 \mbox{$F$ ($=I+J$)} that varies between \mbox{$|\,I-J\,|$} and  \mbox{$I+J$} 
 (Fig.~\ref{fig:energy_diagram} shows an energy diagram). The rotational transition $J$\,=\,1--0 has three HFS lines: \mbox{$F$\,=\,0--1}, \mbox{$F$\,=\,2--1}, and \mbox{$F$\,=\,1--1,} which are separated by \mbox{$-$\,7.1\,km\,s$^{-1}$} and
\mbox{$+$\,4.9\,km\,s$^{-1}$} from the central \mbox{$F$\,=\,2--1} line (see Fig.~\ref{fig:lines_sketch}).
These separations are larger than the typical line widths \mbox{($\sim$0.5--3~km\,s$^{-1}$)} seen at GMC clump
scales \mbox{($\sim$0.2--2~pc)}. Hence,
observations of galactic disk GMCs spectrally resolve these HFS lines (or at least their 
intensity peaks). As in the case of other interstellar  molecules with \mbox{resolved} HFS structure
(N$_2$H$^+$, CF$^+$, NH$_3$,  C$^{17}$O, OH, ...),
 the relative \mbox{$J$\,=\,1--0} HFS line intensity ratios can provide
 straightforward  information
 on the excitation temperature ($T_{\rm ex}$) and column density \citep[e.g.,][]{Caselli02,Guzman12,Punanova18,Zhang2020}. 
This is particularly useful as widefield emission maps
usually detect a single rotational line (often the \mbox{$J$\,=\,1--0}).
However, as we show later, detailed excitation models are needed when
 radiative effects become important.

In the optically thin  limit ($\tau$\,$\rightarrow$\,0), the relative strengths of HCN \mbox{$J$\,=\,1--0} HFS lines
are 1:5:3, $T_{\rm ex}$ is exactly the same for the three HFS transitions and equal to $T_k$ if  local thermodynamic equilibrium (LTE) prevails. \mbox{Therefore}, we would expect the integrated line intensity ratios 
to vary from  \mbox{$R_{12}$\,=\,$W$($F$=1--1)\,/\,$W$($F$=2--1)\,=\,3/5\,=\,0.6}
and \mbox{$R_{02}$\,=\,$W$($F$=0--1)\,/\,$W$($F$=2--1)\,=\,1/5\,=\,0.2}, to \mbox{$R_{12}$\,$=$\,1} and \mbox{$R_{02}$\,$=$\,1}
in the optically thick limit ($\tau$\,$\rightarrow$\,$\infty$). However, observations of interstellar HCN mostly show anomalous HCN 
 ratios, that is, $R_{12}$ and $R_{02}$ values  that are out of the \mbox{[0.6$-$1]} and \mbox{[0.2$-$1]} ranges.

Early observations of warm GMCs detected anomalous \mbox{$R_{12}$\,$<$\,0.6} and \mbox{$R_{02}$\,$\gtrsim$\,0.2} ratios \citep[][]{Gottlieb75,Baudry80}, whereas 
cores in nearby cold dark clouds such as Taurus  \citep[those forming low-mass stars only,][]{Bergin07} show a  more complicated behavior, reaching  \mbox{$R_{12}$\,$>$\,1} and \mbox{$R_{02}$\,$>$\,1} \citep{Walmsley82}. In  dark clouds,  anomalous ratios prevail over large spatial scales  \citep{Cernicharo84}.
More recently, higher angular resolution observations of larger samples of low- and high-mass star-forming cores confirm
the ubiquity of  anomalous intensity ratios \citep[e.g.,][]{Sohn07,Loughnane12,Magalhes:18}.
 As we show later in this work, anomalous HFS ratios immediately imply
that a single $T_{\rm ex}$ cannot describe the excitation of the three \mbox{$J$\,=\,1--0} HFS 
lines\footnote{This is exactly the basic assumption of automatic HFS gaussian  line-fitting programs: same $T_{\rm ex}$ and same line width for all HFS lines.}.

For higher $J$ rotational levels, HCN HFS lines  get closer and, owing
to  bulk gas motions  as well as  to thermal and turbulent  line broadening, most of them overlap \mbox{(colored arrows in Fig.~\ref{fig:lines_sketch})}.
As line opacities rise,  these overlaps induce \mbox{photon} exchanges between  different HFS levels and alter their populations. This  leads to
anomalous HFS  intensity ratios \citep[e.g.,][]{Guilloteau81,Gonzalez93,Turner97}.
Still, most popular large velocity gradient (LVG) radiative transfer codes do not treat HFS line overlaps. While this simplifying approach leads to reasonable results, it nonetheless  $i)$ misses the diagnostic power of HFS lines  and $ii)$ can easily  lead to wrong
 abundances and overestimate the gas density derived from  HCN, HNC, or N$_2$H$^+$
observations \citep[e.g.,][]{Daniel08}.

In this paper, we reassess the role of  HFS line overlaps in the HCN emission from GMCs.
We first compute new \mbox{HFS-resolved} rate coefficients for inelastic collisions of HCN with \mbox{ortho-H$_2$ ($I=1$)} and 
 \mbox{para-H$_2$ ($I=0$)} at gas temperatures higher than computed before \citep[$T_k$\,$\leq$\,30\,K for collisions with \mbox{para-H$_2$};][]{Magalhes:18}.  We also study the \mbox{HCN $J$\,=\,1--0} HFS anomalies in conditions relevant
to the more translucent and extended GMC cloud environment that are not, thus, directly associated with dense star-forming gas. 
In particular, we  investigate the role of \mbox{HFS-resolved} electron collisions
and of gas velocity dispersion (line widths)  in the $R_{02}$ and $R_{12}$ intensity ratios.

The paper is organized  as follows: In \mbox{Sect.~\ref{sec:RT}} we briefly review the basic formalism 
we implemented to treat line overlap in our Monte Carlo (MTC) radiative transfer code.
In \mbox{Sect.~\ref{sec:colisional_rates}} we present the inelastic collisional rate coefficients we use in the models. We present our model results  in {Sect.~\ref{Sec-models}}.
Finally, in \mbox{Sec.~\ref{Sec-applications},} we apply  our models
to the  anomalous \mbox{HCN\,$J$\,=\,1--0} HFS spectra observed in the
Orion Bar and the Horsehead PDRs and to the low-surface-brightness HCN emission  GMCs.

\section{Background: Transfer of overlapping lines}\label{sec:RT}

The main difference compared to the excitation of \mbox{non-overlapping} lines is that
a photon 
 emitted in the  HFS transition known as ``$J_{F}$'', \mbox{meaning} 
\mbox{($J_u,F_u$)\,$\rightarrow$\,($J_u-1,F_l$)}, with a frequency of \mbox{$\nu$\,=\,$\nu_{ul}$($J_{F}$)\,+\,$\delta \nu$},
  can stimulate emissions and absorptions in a different HFS  transition, namely "$J_{F'}$," 
if their line profiles overlap\footnote{Previous theoretical studies of HCN line overlaps 
include simple \mbox{\textit{local}} escape
probability and LVG models  \citep{Gottlieb75,Guilloteau81,Zinchenko87,Turner97},
accurate
\mbox{\textit{nonlocal}} Monte Carlo and \mbox{$\Lambda$-iteration} models 
 \citep{Lapinov89,Gonzalez93} and   \mbox{accelerated}-convergence models  \citep{Daniel08,Mullins16}.}.
Red, green, and cyan colored arrows in  Fig.~\ref{fig:lines_sketch} show the relevant
HFS overlapping lines in  \mbox{HCN $J$\,=\,2-1} and \mbox{3-2} transitions. 
 Photon exchanges between these lines modify the HFS level populations and  the resulting mm-wave HCN spectrum compared to the case in which these exchanges are not considered. 
 As we demonstrate here,  line overlap effects result 
 in anomalous intensity ratios for a broad range of physical conditions.
  In the optically thin limit 
\mbox{($\tau_{ul} \rightarrow 0$)}, photon exchanges between overlapping lines tend to zero (e.g., HFS lines of the rare isotopologues H$^{13}$C$^{14}$N and H$^{12}$C$^{15}$N). However, they become very important as  line opacity increases (\mbox{i.e., HFS lines} of the more abundant species \mbox{HCN\,$=$\,H$^{12}$C$^{14}$N}). For overlapping HFS lines,
the total gas (line) plus dust (continuum) absorption coefficient, \mbox{$\alpha_{v}$\,=\,$\alpha_{\rm d}$\,+\,$\alpha_{\rm g}$}  $[$cm$^{-1}$$]$, at a given velocity $v$ of the  \mbox{$J\rightarrow J-1$} rotational line profile is:
\begin{equation}
\label{eq:abs}
\alpha_{v} =  \alpha_{\rm d} + \sum_{ul=J_F} \alpha_{v,\,ul}  =  
  \alpha_{\rm d} + \sum_{ul=J_F} \frac{hc}{4\pi}
\left( n_l\,B_{lu}- n_u\,B_{ul}\right)\, \phi_{lu}(v),
\end{equation}
where $\alpha_{\rm d}$ refers to the dust absorption coefficient and
the sum extends to all HFS $J_F$ lines of the \mbox{$J\rightarrow J-1$} transition. In this expression, $n_{u}$ and $n_{l}$ refer to the upper and lower HFS level populations in cm$^{-3}$,  $B_{lu}$ and $B_{ul}$ are the Einstein coefficients for stimulated absorption and emission respectively, and $\phi$ is the line profile of each  HFS line in the Doppler velocity space (we 
assume $\phi_{lu} = \phi_{ul}$). In this notation, one can express the frequency of a HFS transition ($\nu_{ul}$)  in a group of overlapping HFS lines in terms of a relative velocity $v_{ul,r}$  [km\,s$^{-1}$] of the $J\rightarrow J-1$ profile: 
 \begin{equation}
\label{eq:frequencies}
v_{ul,r} = \left(1 - \nu_{ul}/\nu_r \right)\,c,
\end{equation}
 where $\nu_r$ is a reference frequency.  Here, we choose that of the strongest 
(in intrinsic line strength) HFS component of each  \mbox{$J\rightarrow J-1$}  transition (see  Fig.~\ref{fig:lines_sketch}).
 The total opacity at $v$ is then
$\tau_{v} = \alpha_{v}\,\Delta x$, with $\Delta x$ in cm.

To model line overlap effects we modified  a  multi-slab   MTC code\footnote{Since we are mostly interested in the extended GMC  emission, where HCN line opacities are expected to be \mbox{$\tau_{\rm hfs}$\,$<$\,100}, we do not attempt to implement any convergence acceleration method.}
that treats both spherical and plane-parallel geometries 
\citep[Appendix of][]{Goicoechea_2006}.
  With some adjustments  to the original code \citep[][]{Bernes79}, 
  one can compute
the number of stimulated emissions, \mbox{$s$($J_{F'}$,$J_{F}$)}, that  
a model photon representing a number of real line photons emitted in the
 HFS transition, $J_{F}$, produces, as it travels through the cloud, in all overlapping transitions, $J_{F'}$.
We successfully benchmarked our procedure and model results against
the HCN HFS test problem M$_{401}$ of \cite{Gonzalez93}.

In the \mbox{classic} \mbox{Monte Carlo} method for non-overlapping lines,  a  photon emitted in the  $J_{F}$ line only produces stimulated emissions in this same transition. That is to say,
\mbox{$B_{ul}$$\bar{J}_{ul}$\,$\propto$\,$s_{ul}$($J_{F}$)} (in Bernes formalism), where
$\bar{J}_{ul}$ is the  mean intensity of the $ul$ line at a given position of the cloud
(that in turn  depends, non-locally, on the physical conditions and HFS level populations in  other cloud positions). With line overlap, the number of induced stimulated emissions in the  $J_{F'}$ transition at a given cloud position  is:
\begin{equation}
\label{eq:overlap}
s_{u'l'}(J_F')=\sum_{ul=J_F} s(J_{F'},J_{F})
,\end{equation}
where the  sum   includes all overlapping HFS lines. This is a more time consuming calculation because one has to follow each line photon and compute
the number of stimulated emissions that it causes in all  overlapping transitions \citep[see also][]{Lapinov89,Gonzalez93}. The same applies to continuum photons emitted
by dust grains and the cosmic microwave background.
They can now be absorbed by different HFS overlapping lines $J_{F'}$, depending
on their line opacity ratio
$\tau_{J_{F'}}/\sum{\tau_{J_{F}}}$ at each velocity position $v$ in the line profile. 

With line overlap, the velocity-dependent source function is:
\begin{equation}
\label{eq:source}
S_{v} = \frac{j_{v}}{\alpha_{v}} = 
\frac{j_{\rm d} + j_{\rm g} }{\alpha_{\rm d} + \alpha_{\rm g}} = 
 \frac{j_{\rm d} + \sum_{ul} j_{v,\,ul}} {\alpha_{\rm d} + \sum_{ul} \alpha_{v,\,ul}} \,,
\end{equation}
where $j_{\rm d}$ and $j_{\rm g}$ are the dust and HFS line emissivity coefficients:
\begin{equation}
\label{eq:emissivity}
j_{\rm d} = \alpha_{\rm d}\, B\,(T_{\rm d})\,\,\,\, ; \,\,\,\, 
j_{v} = \frac{hc}{4\pi}\, n_u\, A_{ul}\,\phi_{ul} (v) \,,
\end{equation} 
 where $B$ is the Planck function at a dust grain temperature, $T_{\rm d}$, and $A_{ul}$ is the Einstein coefficient 
for spontaneous emission of HFS transition, $ul$. In most interstellar applications,
the low-lying rotational  lines of HCN are not affected by dust opacity 
(\mbox{i.e., $\alpha_{\rm d}\rightarrow 0$}). In this case,
we can simply write the source function as:
\begin{equation}
\label{eq:source}
S_{v} = \frac{1}{\alpha_g} \sum_{ul} \alpha_{v,\,ul}\,\,S_{ul} \,,
\end{equation}
where $S_{ul}$ is the standard
velocity-independent source function, \mbox{$B$\,($\nu_{ul}$, $T_{{\rm{ex}},\,ul}$)}, of each individual HFS line.

After a Monte Carlo simulation of the line and continuum photons, we determine the HFS level populations  by solving
the steady-state statistical equilibrium equations:
\begin{equation}
n_u \sum_{l\neq u} \left( R_{ul} + C_{ul} \right) =
\sum_{l\neq u} n_l\,\left( R_{lu} + C_{lu} \right),
\end{equation}
where $C_{ul}$ and $R_{ul}$  are the collisional and radiative pumping rates
[s$^{-1}$] between the HFS levels $u$ and $l$. The radiative rates are:
\begin{equation}
R_{ul}=A_{ul}\,+\,B_{ul}\,\bar{J}_{ul} \,\,\,\,\,;\,\,\,\,\,\,
R_{lu} = B_{lu}\,\bar{J}_{lu}  \,\,\,\,\,;\,\,\,\,\,\, B_{ul}\,\bar{J}_{ul} = \sum_{} s_{ul} \,,
\end{equation}
with $s_{ul}$ as defined in Eq.~\ref{eq:overlap}.
For the collisional rates, we consider HFS-resolved collisions of HCN  molecules with \mbox{ortho-H$_2$}, \mbox{para-H$_2$}, and electrons:
\begin{equation}
C_{ul} = k_{ul}\,(o{\rm -H_2})\,n\,(o{\rm -H_2}) 
       + k_{ul}\,(p{\rm -H_2})\,n\,(p{\rm -H_2})
       + k_{ul}\,({\rm e^-})\,n\,({\rm e^-})\,,
\end{equation}
where $n$ [cm$^{-3}$] refers to the density of each collisional \mbox{partner} \mbox{(\mbox{ortho-H$_2$}, \mbox{para-H$_2$}, and $e^-$)} and $k_{ul}$ [cm$^{3}$\,s$^{-1}$] are the temperature-dependent  collisional
de-excitation rate coefficients. These are obtained after detailed quantum calculations from thermal averages of the specific
collision cross-sections \mbox{\citep[e.g.,][]{Roueff13}}.  In addition to a correct treatment of
 line overlaps, these  coefficients are the critical ingredients to accurately determine the excitation and line emission from any interstellar molecule.
In the following section, we summarize the  rate
coefficients \mbox{$k_{ul}$($T_{\rm k}$)}  that we adopted in this study.

\section{Collisional excitation of HCN HFS levels}\label{sec:colisional_rates}

\subsection{New inelastic collisions with ortho- and para-H$_2$}

\cite{Monteiro:86} calculated the first HFS-resolved \mbox{HCN-He} 
 quantum  collisional rate coefficients  for rotational levels \mbox{$J$\,=\,0 to 4}  and $T_{\rm k}$ ranging from 10 to 30\,K (with He as a surrogate for \mbox{para-H$_2$}). 
It wasn't until much later that \cite{Abdallah:12}  computed the rate coefficients for the hyperfine (de-)excitation of HCN in collisions with  \mbox{para-H$_2$\,($J_2$\,=\,0);} we note that we use 
$J_2$ here to designate the rotational level of H$_2$ and avoid confusion with the rotational level $J$ of HCN. These authors  used a potential energy surface (PES) averaged over the H$_2$ orientations. Although the work by \cite{Abdallah:12} improves the accuracy of the rate coefficients compared to models using \mbox{HCN-He} collisional rates, these calculations consider H$_2$ as a structureless collisional partner and, as discussed by \cite{Vera:14}, this approximation leads to significant inaccuracies. To overcome this issue, \mbox{\cite{Magalhes:18}} calculated new 
 \mbox{HCN\,--\,$p$-H$_2$\,($J_2$=\,0)} HFS-resolved rate coefficients, for the lowest 25 hyperfine levels and $T_{\rm k}$ in the range \mbox{5--30\,K},
using the nearly exact \mbox{recoupling} method \citep{Lanza:14}. These data use a full-dimension PES \citep{Denis:13} and are very accurate. Unfortunately, their utility is limited to low-temperature environments 
\mbox{(e.g., cold dark clouds and prestellar cores)}. 
This lack of HFS-\mbox{resolved} rates for warm gas applications motivated us to
 calculate the rate coefficients for HCN collisions with both \mbox{para-H$_2$($J_2$=0)} and     \mbox{ortho-H$_2$($J_2$=1)},  up to \mbox{$J$\,=\,11} \mbox{(34 HFS levels)}
 and  \mbox{$T_{\rm k}$\,$\leq$\,500\,K}. However, because of the small rotational constant of HCN and the high temperatures targeted, recoupling calculations are not feasible.  Instead, we determine the  HFS-resolved rate coefficients, 
 using the 
 scaled-infinite order sudden limit \mbox{(S-IOS)} \mbox{approximation}, 
  from the highly accurate pure rotational rate coefficients 
obtained by \cite{Vera:17} in the exact \mbox{close coupling}  method (CC).  The
\mbox{IOS} approach was first introduced by \cite{Neufeld:94} for diatom--atom collisions and later extended
 by \cite{Lanza:14}
  to the HFS excitation of linear molecules in collisions with \mbox{para-H$_2$} and \mbox{ortho-H$_2$}. In practice, we scaled the CC rotational rate coefficients with the ratio of HFS and rotational IOS rate coefficients as:
\begin{equation} \label{formula_SIOS}
k^{S-IOS}_{J,F,\,J_2 \to J',F',\,J_2'}=
\frac{k^{IOS}_{J,F,\,J_2 \to J',F',\,J_2'}}
{k^{IOS}_{J,\,J_2 \to J',\,J_2'}} k^{CC}_{J,\,J_2 \to J',\,J_2'}\,\,,
\end{equation}
where the methodology  to calculate $k^{IOS}_{J,F,\,J_2 \to J',F',\,J_2'}$ and 
$k^{IOS}_{J,\,J_2 \to J',\,J_2'}$ is described by \cite{Lanza:14}.
\cite{Faure:16} previously applied this approach to the \mbox{HC$_3$N-H$_2$} system and
found  that the \mbox{S-IOS} approximation is very accurate for collisions with 
\mbox{para-H$_2$($J_2$\,=\,0)}, and it predicts the correct behavior at intermediate and high kinetic energies in collisions with 
\mbox{ortho-H$_2$($J_2$\,=\,1)}. In both cases, the agreement increases with increasing 
$T_{\rm k}$. 

\begin{figure*}[t]
\centering   
\includegraphics[scale=0.65, angle=0]{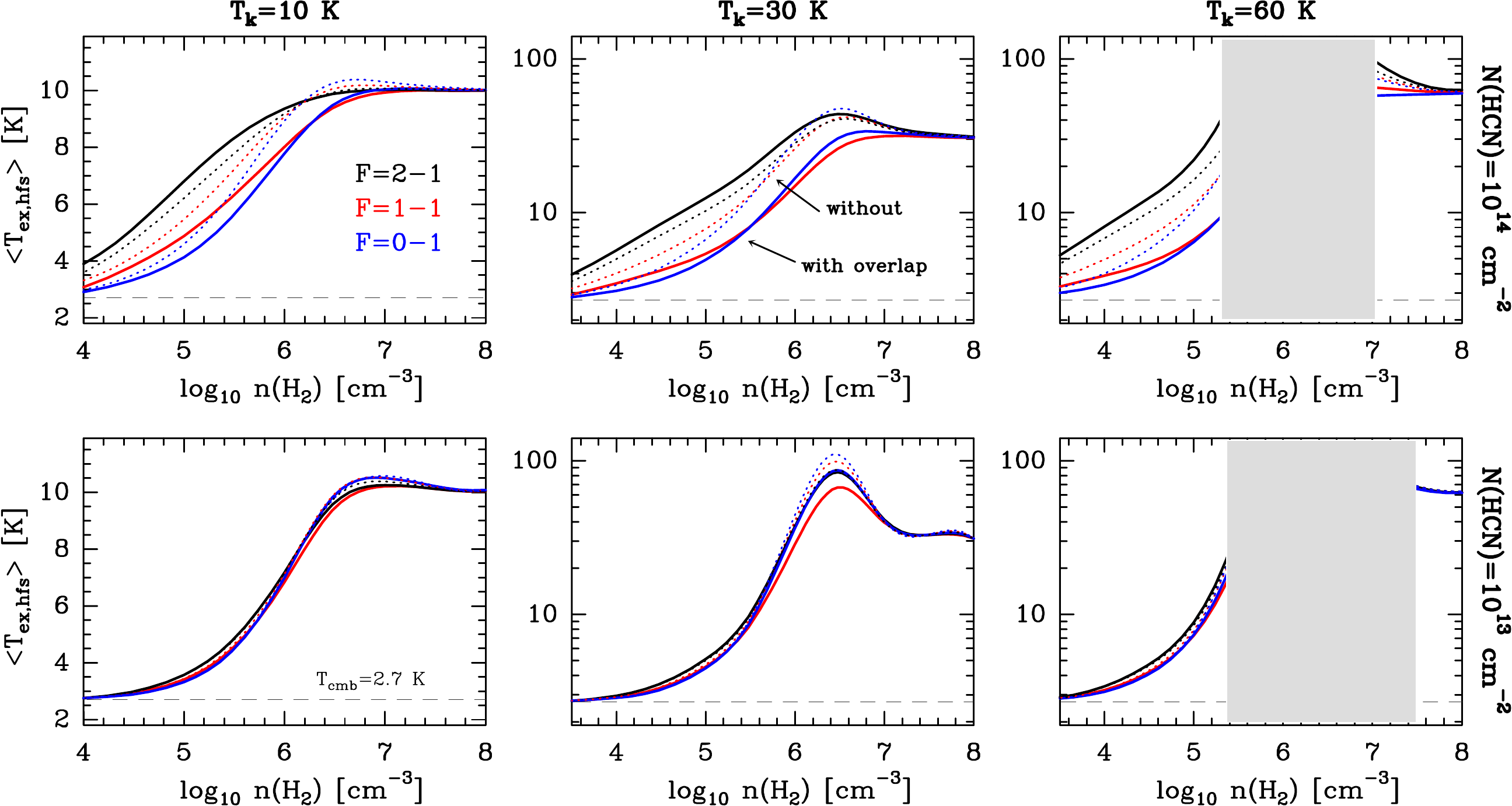}
\caption{Mean excitation temperature \mbox{$T_{\rm ex}$($F_{\rm u}$--$F_{\rm l}$)} of the three  \mbox{HCN $J$\,=\,1--0} HFS lines obtained from nonlocal  non-LTE  models of constant density and temperature clouds with \mbox{$\Delta v_{\rm turb}$\,=\,1\,km\,s$^{-1}$}. $N$(HCN) is \mbox{10$^{14}$\,cm$^{-2}$} in the upper panels
and \mbox{10$^{13}$\,cm$^{-2}$} in the lower panels. Dotted curves show  models that
neglect  line overlaps between different HFS lines.
The gray-shaded areas are regions of the parameter space where the excitation is very suprathermal
(\mbox{$T_{\rm ex}$\,$\gg$\,$T_{\rm k}$}) or weakly masing (\mbox{$T_{\rm ex}$\,$<$\,0
with small \mbox{$|\,\tau\,|$} )}.} 
\label{fig:grid_Tex}
\end{figure*}

 In order to evaluate the accuracy of this method, we compared the
 \mbox{HCN-p-H$_2$} rate coefficients obtained in the \mbox{S-IOS} approximation
 at temperatures below 30\,K   with those of \cite{Magalhes:18}. The agreement between the two data sets is better than 10-20\% for the dominant rate coefficients, those corresponding to transitions with small \mbox{$\Delta J$ (\,=\,1, 2, ...)} or with 
 $\Delta J=\Delta F$, and better than 30-50\% for the others. Because the validity domain of IOS based methods increases with increasing temperature, we expect the accuracy of the
 new rate coefficients to be better than 10-20\% for temperatures above 50\,K.
 The dataset also includes quasi-elastic (\mbox{$\Delta J$\,=\,0}) collisional rate coefficients, computed from a pure IOS approach because CC elastic rate coefficients
are not available.
 \mbox{Quasi-elastic}  collisions mix  HFS level populations. Thus,
 as first discussed by \cite{Guilloteau81} and \cite{Gonzalez93},
 the absolute value of these quasi-elastic rates  influences the resulting line intensities and the degree of anomalous intensity ratios.
\mbox{Although} the low-temperature quasi-elastic collision rate coefficients calculated by \mbox{\cite{Magalhes:18} }  
are more accurate, we checked that
 radiative transfer models using those computed in the \mbox{S-IOS} \mbox{approximation} 
 result in comparable line intensities and  similar $R_{12}$ and $R_{02}$ intensity ratios
 (see Appendix~\ref{App-recoupling-IOS}). 

Here, we adopt HFS-resolved \mbox{HCN\,--\,$p$-H$_2$} collisional rate coefficients
computed in the recoupling method for models with \mbox{$T_{\rm k}$\,$\leq$\,30\,K} (at such low temperatures the abundance of \mbox{ortho-H$_2$} is low, hence collisions with \mbox{ortho-H$_2$} do not play a role), and those  computed in the \mbox{S-IOS} approximation
for \mbox{HCN\,--\,$o$-/$p$-H$_2$} collisions at  \mbox{$T_{\rm k}$\,$>$\,30\,K}. 
We stress that until 2016, all  studies of HCN HFS anomalies
 used scaled \mbox{HCN--He}  rate coefficients and approximated
 quasi-elastic rates.  However, the standard reduced-mass scaling
  commonly
used to estimate \mbox{$k_{ul}$(H$_2$)} from \mbox{$k_{ul}$(He)}
(multiplying by 1.37)  is often a crude estimation, particularly when collisions with ortho-H$_2$ are relevant 
\citep[see also][]{Walker14}.

 Here we follow a more accurate treatment and explicitly account for collisions with both
  \mbox{para-H$_2$} and \mbox{ortho-H$_2$} in warm gas. We assume that the H$_2$ \mbox{ortho-to-para}  ratio (OPR), 
\mbox{$n$($o$-H$_2$)/$n$($p$-H$_2$)}, is thermalized to the gas temperature, for instance:~\mbox{OPR($T_{\rm k}$=30\,K)\,$\simeq$\,0.03}, \mbox{OPR($T_{\rm k}$=60\,K)\,$\simeq$\,0.5}, \mbox{OPR($T_{\rm k}$=100\,K)\,$\simeq$\,1.6},
and  \mbox{OPR($T_{\rm k}$=200\,K)\,$\simeq$\,2.8}. We note, \mbox{however}, that particular environments
such as protostellar shocks and PDRs may have H$_2$ OPRs that deviate from 
exact thermalization \mbox{\citep[e.g.,][]{Neufeld06,Habart11}}.
As an \mbox{example}, Table~\ref{rates} shows de-excitation
rate coefficients\footnote{The full dataset of rate coefficients will be available online.} for the \mbox{HCN $J$\,=\,1\,$\rightarrow$\,0 ($F$\,=\,2\,$\rightarrow$\,1)} HFS transition calculated
in the \mbox{S-IOS} approximation at different gas temperatures and OPRs.

\begin{table}[t]
  \begin{center}
    \caption{\label{rates} Collisional de-excitation rate coefficients 
    for the \mbox{HCN $J$\,=\,1\,$\rightarrow$\,0 ($F$\,=\,2\,$\rightarrow$\,1)} transition (in powers of 10 and \mbox{cm$^3$\,s$^{-1}$}).}
    \begin{tabular}{rrrr}
      \hline \hline
       $T_{\rm k}$        &  10\,K        &  50\,K      & 200\,K\\
 H$_2$ OPR                &  $\sim$0.0    &   0.3       & 2.8    \\      
      \hline
     $k$($p$-H$_2$)${^\dagger}$  &  2.46($-$11)  & 2.23($-$11) & 2.11($-$11)\\
     $k$($o$-H$_2$)${^\dagger}$  &  1.66($-$10)  & 1.58($-$10) & 1.67($-$10)\\
     $k$(H$_2$ OPR)                              &  2.46E($-$11) & 5.34($-$11) & 1.29($-$10)\\
     $k$(e$^-$)${^\ddagger}$     &  3.73($-$6)   & 3.29($-$6)  & 2.61($-$6)\\
      \hline
      \end{tabular}
   \tablefoot{${^\dagger}$This work. ${^\ddagger}$From \cite{Faure07}.}
  \end{center}
\end{table}

\begin{table}[t]
  \begin{center}
    \caption{\label{ncr} Critical densities and critical ortho-H$_2$ and electron fractional
    abundances for the \mbox{HCN $J$\,=\,1\,$\rightarrow$\,0 ($F$\,=\,2\,$\rightarrow$\,1)} transition.}
    \begin{tabular}{rrrr}
      \hline \hline
       $T_{\rm k}$               &  10\,K        &  50\,K      & 200\,K\\
 H$_2$ OPR                       &  $\sim$0.0    &   0.3       & 2.8    \\      
      \hline
 $n_{\rm cr}$(H$_2$) [cm$^{-3}$] &  9.80($+$5)   & 4.50($+$5)  & 1.87($+$5) \\
$\chi^{*}_{\rm cr}$($o$-H$_2$)   &  0.15  & 0.14 & 0.13 \\
 $n_{\rm cr}$(e$^-$) [cm$^{-3}$] &  6.45         & 7.31        & 9.22 \\
$\chi^{*}_{\rm cr}$(e$^-$)       &  6.58($-$6)  & 1.62($-$5)   & 4.93($-$5) \\
      \hline
      \end{tabular}
  \end{center}
\end{table}

\subsection{Inelastic collisions with electrons}\label{sec:elec_col}

It has long been recognized that electron collisions  contribute to the rotational excitation of HCN  
 \citep[e.g.,][]{Dickinson77} in  environments where, first,  
the H$_2$ density is not too high to thermalize a given transition; \mbox{$n$(H$_2$) less than several 10$^{5}$\,cm$^{-3}$} for  \mbox{HCN\,$J$\,=\,1--0} 
\mbox{\citep[e.g.,][]{Goldsmith17}} and, second, the
ionization fraction is high, $\chi_{\rm e}$\,$\geq$\,10$^{-5}$.
That is,
greater than the inelastic-collision rate-coefficient ratio \mbox{$k_{ul}$(HCN--H$_2$)\,/\,$k_{ul}$(HCN--e$^-$)}.
Here, we investigate the role of electron collisions in the HCN HFS anomalies using
specific HFS-resolved \mbox{$k$(HCN--e$^-$)} rate coefficients 
calculated by \cite{Faure07}  using the IOS scaling of the \mbox{$k_{J \to J'}$} rates  for $J$\,$\leq$\,8 and electron temperatures ($T_{\rm e}$) in the range \mbox{5--6000\,K}.
Table~\ref{rates} lists the   
\mbox{HCN $J$\,=\,1\,$\rightarrow$\,0 ($F$\,=\,2\,$\rightarrow$\,1)} 
de-excitation rate coefficients.
Contrary to H$_2$ collisions, electron collisions favor the  $\Delta J$\,=\,1 (dipole-like) transitions, with a strong propensity
rule $\Delta J$\,=\,$\Delta F$ \citep{Faure07}.
In our models  (collisions with both H$_2$ and electrons), 
we calculate the 
corresponding collisional excitation rates  assuming detailed balance and $T_{\rm k}$\,=\,$T_{\rm e}$ (thermalization).

\subsection{Critical densities and critical fractional abundances}\label{sec:crit_dens}

The density at which the collisional \mbox{de-excitation} rate  equals the spontaneous emission rate is often referred to as
the \mbox{``critical density''} of a given transition. For a two-level system:
\begin{equation} \label{formula_ncr}
n_{\rm cr}\,({\rm{H_2\,\,or\,\,e^-}}) = \frac{A_{ul}}{k_{ul}(T_{\rm k})}\,\,[{\rm cm^{-3}}]
.\end{equation}
Table~\ref{ncr} shows \mbox{$n_{cr}$(HCN $J$\,=\,1\,$\rightarrow$\,0 $F$\,=\,2\,$\rightarrow$\,1)} for different
collisional partners and temperatures. In the weak collisional excitation limit 
($n \ll n_{\rm cr}$) radiative effects dominate
the excitation of a given transition $ul$ and \mbox{$T_{{\rm ex},\,ul}$ tends to  
$T_{\rm cmb}$\,=\,2.7\,K}. 
In the strong collisional limit ($n \gg n_{\rm cr}$), collisions drive the excitation toward LTE, with
 \mbox{$T_{\rm ex}\simeq T_{\rm k}$}, and $T_{\rm ex}$ is the same excitation temperature for  all HFS transitions. In practice, as line opacity $\tau_{ul}$ increases, line-trapping reduces the ``effective'' critical
 density, roughly as $n_{\rm cr,\,eff} \simeq n_{\rm cr}/\tau_{ul}$ 
 \mbox{\citep[e.g.,][]{Shirley15}}.
 For rotationally excited lines, the critical densities \mbox{$n_{\rm cr}$($J\rightarrow J-1$)} of the \mbox{higher-$J$} HCN lines quickly increases,
 with \mbox{$n_{{\rm cr},\,J=3-2}$\,$\simeq$\,30\,$n_{{\rm cr},\,J=1-0}$}.

It is also useful to define the critical electron fractional abundance, \mbox{$\chi_{\rm cr}^{*}$(e$^-$)}, at
which the electron collision rate equals the H$_2$  rate of a given transition \citep[e.g.,][]{Goldsmith17}. For collisions with both \mbox{ortho-H$_2$} and \mbox{para-H$_2$}, this 
implies:
\begin{equation}\label{formula_xcr_e}
\chi_{\rm cr}^{*}({\rm e^-}) = 
\frac{\frac{{\rm OPR}}{1+{\rm OPR}}\, k_{ul}({\rm {\it o}{\rm -}H_2}) + 
 \frac{1}{1+{\rm OPR}}\, k_{ul}({\rm {\it p}{\rm -}H_2})}{k_{ul}(e^-)}.
\end{equation}
Although $\chi_{\rm cr}^{*}({\rm e^-})$ slightly varies with temperature,
electron collisions start to dominate the excitation of 
\mbox{HCN $J$\,=\,1--0} HFS lines at ionization fractions of about 
\mbox{$\geq$\,10$^{-5}$} if 
\mbox{$n$(H$_2$)\,$<$\,$n_{\rm cr}$(H$_2$), as shown in} \mbox{Table~\ref{ncr}}. Likewise, we can define the critical ortho-H$_2$ fractional abundance, 
\mbox{$\chi_{\rm cr}^{*}$($o$-H$_2$)}
at which 
the \mbox{$o$-H$_2$} collision rate equals the \mbox{$p$-H$_2$} collision rate:
\begin{equation}\label{formula_xcr_oH2}
\chi_{\rm cr}^{*}({\rm {\it o}{\rm -}H_2}) =
 \frac{k_{ul}({\rm {\it p}{\rm -}H_2})}{k_{ul}({\rm {\it o}{\rm -}H_2})} =
 \frac{n_{\rm cr}({\rm {\it o}{\rm -}H_2})}{n_{\rm cr}({\rm {\it p}{\rm -}H_2})}.
\end{equation}
 Collisions with $o$-H$_2$ start to dominate for H$_2$ OPR values of $\gtrsim$\,0.15 \mbox{(Table~\ref{ncr})}, which implies gas temperatures of
 \mbox{$T_{\rm k}$\,$>$\,40\,K} if the OPR is thermalized to $T_{\rm k}$.

\begin{figure*}[t]
\centering   
\includegraphics[scale=0.65, angle=0]{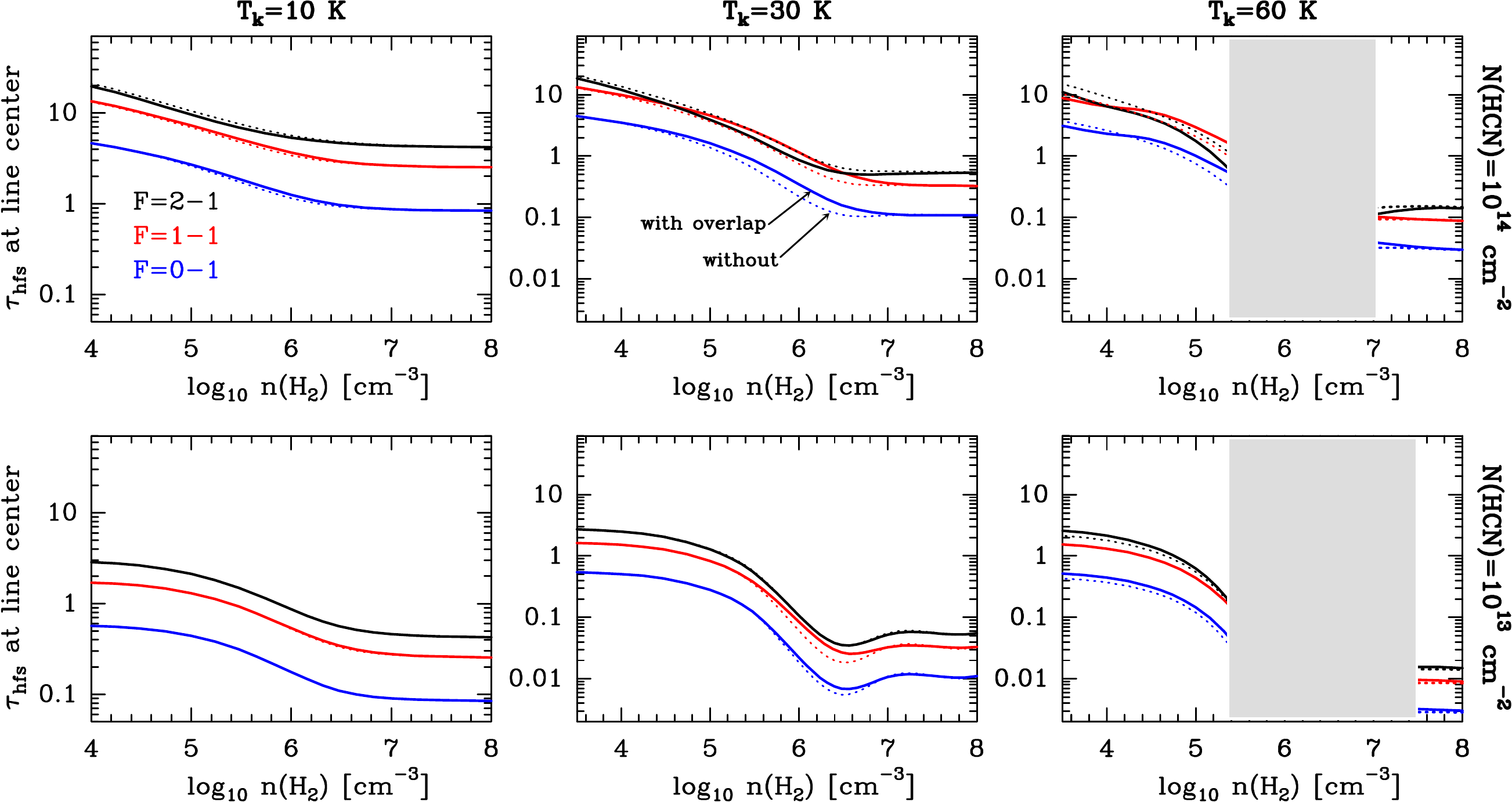}
\caption{\mbox{HCN $J$\,=\,1--0}  HFS line opacities obtained from our grid of models. Same details as in \mbox{Fig.~\ref{fig:grid_Tex}} but for the  opacities at the line center.} 
\label{fig:grid_tau}
\end{figure*}

\section{Grid of HCN HFS line emission models}\label{Sec-models}

Here, we explore the role of line overlap and of the new  collisional rate coefficients
in models that cover a  broad range of \mbox{physical} conditions relevant to the  emission from GMCs. We ran a grid of single-component MTC models 
for HCN column densities that bracket the typical values observed in GMCs:  $N$(HCN)\,=\,10$^{13}$\,cm$^{-2}$, leading to optically thin or marginally thick 
\mbox{$J$\,=\,1--0} HFS lines, and $N$(HCN)\,=\,10$^{14}$\,cm$^{-2}$, leading to optically thick lines. We cover the gas density range from $n$(H$_2$) of several 
\mbox{10$^3$ cm$^{-3}$}, relevant to the most translucent and extended gas component of GMCs, to  \mbox{$n$(H$_2$)\,=\,10$^{8}$\,cm$^{-3}$}, which is only relevant to the inner layers of hot cores or corinos and of protostellar envelopes. There, inelastic collisions drive the excitation close to LTE.
 The resulting HCN line profiles include \mbox{thermal}, \mbox{microturbulent}, and opacity broadening. Velocity gradients \mbox{(infall or outflows)} also affect the HFS emission  and can produce  even more anomalous intensity ratios \citep[for collapsing cold core models, see][]{Gonzalez93,Magalhes:18}.
As we did not model any specific region and because we are interested in the extended GMC  emission, we did not include any cloud velocity profile.  
Specifically, we ran spherical cloud models with uniform gas densities, temperatures (\mbox{$T_{\rm k}$\,=\,10}, 30, and \mbox{60\,K}), and gas velocity dispersions 
(fixed at \mbox{$\sigma_{\rm turb}$\,=\,0.4\,km\,s$^{-1}$}, with \mbox{$\Delta$v$_{\rm turb,\,FWHM}$\,=\,2.355\,$\sigma_{\rm turb}$}). However, as we
use a multi-slab model (discretized in 40 shells) the excitation temperatures of the HCN HFS lines ($T_{\rm ex,\,hfs}$) are not necessarily uniform throughout the cloud (i.e., line excitation conditions do change) because, as opacities increase, line trapping and cloud boundary effects become important. These radiative  effects, more relevant for subthermally excited ($T_{\rm ex,\,hfs}$\,$\ll$\,$T_{\rm k}$) and  optically thick lines,  are not captured by standard LVG models which, in addition, neglect the radiative
coupling between different cloud positions. 

Since we mainly aim to study the spatially resolved  emission
from GMCs, we calculated the HCN column densities, line intensities, and line intensity ratios for a ray that passes through the center of the modeled spherical cloud 
\mbox{(i.e., along a diameter)}.
We note that more specific models of spatially unresolved emitting sources, or sources
with varying physical conditions and abundances (e.g., prestellar cores), will require multiple ray tracing (i.e., involving a range of HCN column densities) and a convolution of the resulting line intensities
with the telescope beam pattern at each line frequency.

\begin{figure*}[t]
\centering   
\includegraphics[scale=0.65, angle=0]{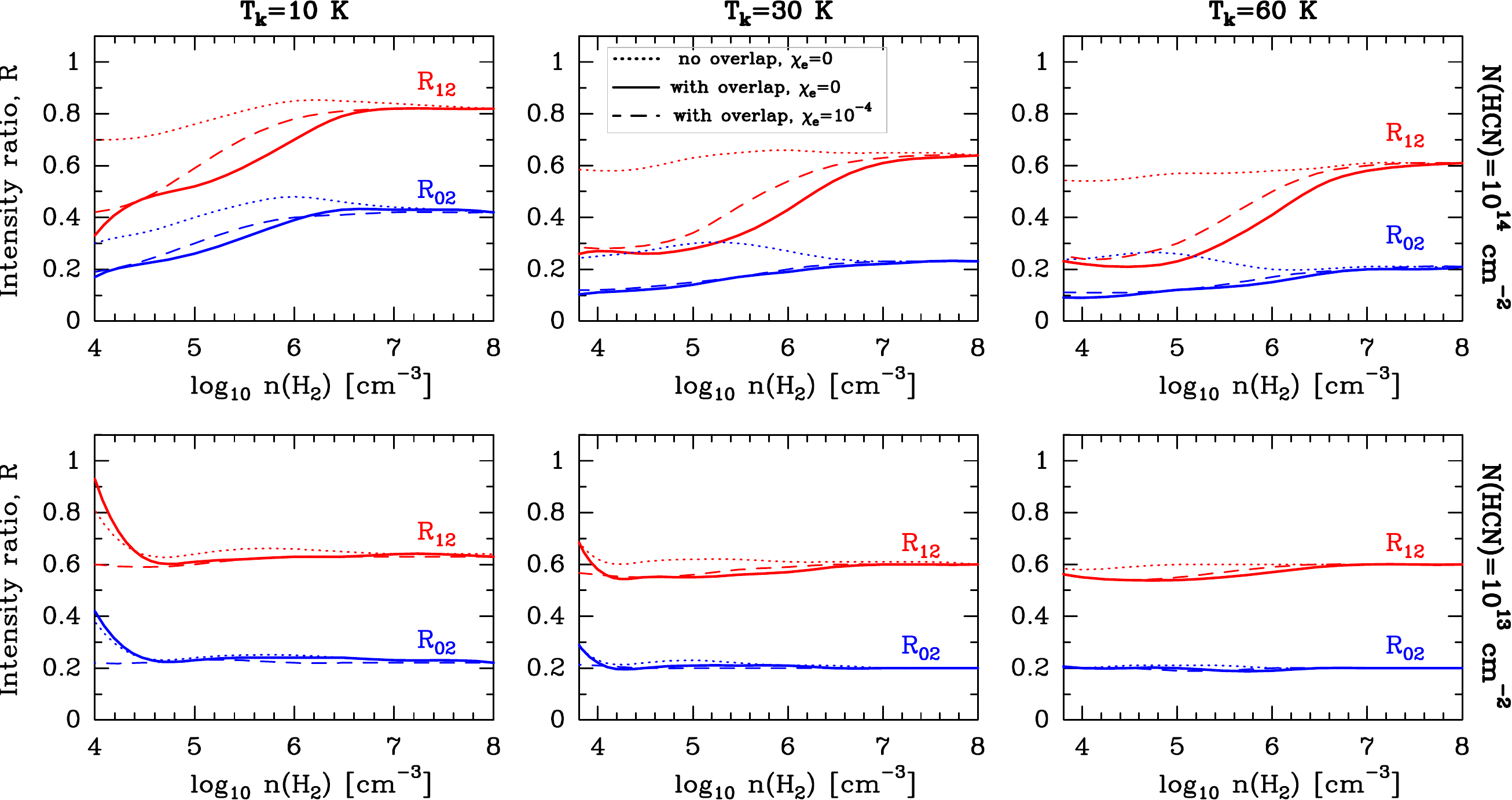}
\caption{\mbox{HCN\,$J$\,=\,1--0}  HFS-integrated line intensity ratios \mbox{$R_{12}$\,=\,$W$($F$=1--1)\,/\,$W$($F$=2--1)}
and \mbox{$R_{02}$\,=\,$W$($F$=0--1)\,/\,$W$($F$=2--1)} obtained from our
grid of models. Same details as in \mbox{Fig.~\ref{fig:grid_Tex}}. 
Dashed curves are models that include line overlap and 
electron excitation with $\chi_{\rm e}$\,=\,10$^{-4}$ (see Sect. \ref{sec:role_elec}).}
\label{fig:grid_R}
\end{figure*}

\subsection{Changes in the HCN $J$=1-0  excitation due to line overlap}
 
We first  investigate the role of HFS line overlap  neglecting electron collisions
(i.e., appropriate to  UV-shielded cloud environments, thus low $\chi_{\rm e}$).
Figure~\ref{fig:grid_Tex} shows the mean excitation temperature of each HFS line ($T_{\rm ex,\,hfs}$ radially averaged over the 40 shells).
Continuous curves show  models that treat line overlap in all considered rotational transitions, whereas the dotted curves refer to models that do not. In the latter case,
we treated the radiative excitation of each HFS line independently
of the others. 
Figure~\ref{fig:grid_tau}  shows the line center opacity of each HFS line ($\tau_{\rm hfs}$).

The different $T_{\rm ex,\,hfs}$ values of each HFS line in Fig.~\ref{fig:grid_Tex} capture the essence of line overlap effects. 
At low line opacities,
photon exchanges between different HFS lines are negligible and line overlap effects
are irrelevant. In addition,  the excitation temperature of the three \mbox{$J$\,=\,1-0} 
 HFS lines is nearly the same in all parameter space.
This is the typical behavior for 
\mbox{$N$(HCN)\,$<$\,10$^{13}$\,cm$^{-2}$}. 
Since it is not easy to collisionally excite a very polar molecule such as HCN, rising $T_{\rm ex,\,hfs}$ to  \mbox{$\gtrsim$\,4\,K} typically requires $n$(H$_2$)  above $\approx$10$^{5}$\,cm$^{-3}$ 
(optically thin gas and \mbox{neglecting} electron excitation).
Beyond that, only at very high densities, $\gtrsim$10$^{7}$\,cm$^{-3}$ (strong collisional limit),
$T_{\rm ex,\,hfs}$ thermalizes to the gas temperature $T_{\rm k}$.

For optically thick HCN lines (\mbox{$\tau_{\rm hfs}$\,$\gtrsim$\,1}) and for densities 
\mbox{$n$(H$_2$)$\,<\,$10$^7$\,cm$^{-3}$}, line-trapping and line-overlap effects
\mbox{alter} the HCN HFS level populations. Even ignoring line \mbox{overlap} effects, the excitation of
 \mbox{HCN $J$\,=\,1--0}  HFS levels is such that
\mbox{$T_{\rm ex,\,{\it F}=\,2-1}$\,$>$\,$T_{\rm ex,\,{\it F}=\,1-1}$\,$>$\,$ T_{\rm ex,\,{\it F}=\,0-1}$} 
(\mbox{dotted curves} in \mbox{Fig.~\ref{fig:grid_Tex}}). 
\cite{Kwan75} first proposed that  these anomalous populations can be explained by the large collisional excitation from \mbox{$J$\,=\,0} to \mbox{$J$\,=\,2} levels
(owing to the high  rates \mbox{$C$($J$\,$=$\,0$\rightarrow$\,2)} in \mbox{HCN--H$_2$ collisions})
 followed by fast radiative decay to \mbox{$J$\,=\,1}  as the   
\mbox{$J$\,=\,2--1} HFS lines become optically thick. In this case, the net
rate of decay from \mbox{$J$\,=\,2$\rightarrow$1} is independent of the line strengths.
For the specific range of H$_2$ densities
\mbox{$\approx$\,10$^{6}$\,cm$^{-3}$} to \mbox{$\approx$\,10$^{7}$\,cm$^{-3}$},
line-trapping in the \mbox{$J$\,=\,2--1} lines reduces the population of the \mbox{$J$\,$=$\,0} level and produce suprathermal \mbox{$J$\,=\,1--0} HFS emission \mbox{($T_{\rm ex}$\,$>$\,$T_{\rm k}$)}.

Line overlaps in the \mbox{$J$\,=2--1} and \mbox{$J$\,=3--2} HFS transitions (colored arrows in Figs.~\ref{fig:lines_sketch} and \ref{fig:energy_diagram}) lead to an increasingly efficient transfer of population from level  \mbox{$J$\,=\,1, $F$\,=\,1} to level \mbox{$J$\,=\,1, $F$\,=\,2} \citep[][]{Guilloteau81,Gonzalez93}. This transfer results in increased $T_{\rm ex,\,{\it F}=\,2-1}$ and decreased  
 $T_{\rm ex,\,{\it F}=\,1-1}$ and $T_{\rm ex,\,{\it F}=\,0-1}$ 
 (continuous curves in \mbox{Fig.~\ref{fig:grid_Tex}}) compared to models that neglect line overlaps \mbox{(dotted curves)}.
For \mbox{$N$(HCN)\,=\,10$^{14}$\,cm$^{-2}$}, \mbox{$T_{\rm k}$\,=\,30\,K}, and
\mbox{$n$(H$_2$)\,=\,10$^5$\,cm$^{-3}$}, the total line opacity at the center of the overlapping groups
\mbox{$J$\,=\,2--1, $F$\,=\,3--2 and $F$\,=\,2--1}  (red arrows) and \mbox{$J$\,=\,3--2, $F$\,=\,4--3, $F$\,=\,3--2, and $F$\,=\,2--1}
(cyan arrows) is \mbox{$\tau_{J=2-1}$\,$\simeq$\,13} and \mbox{$\tau_{J=3-2}$\,$\simeq$\,12}, respectively. 
For \mbox{$N$(HCN)\,=\,10$^{14}$\,cm$^{-2}$} (optically thick \mbox{$J$\,=\,2--1} lines), we predict very suprathermal emission \mbox{($T_{\rm ex}$\,$\gg$\,$T_{\rm k}$)} 
at \mbox{$T_{\rm k}$\,$=$\,60\,K} and  \mbox{$n$(H$_2$)\,$\simeq$\,10$^{5.5}$} to $\simeq$\,10$^{7}$\,cm$^{-3}$.
For \mbox{$N$(HCN)\,=\,10$^{13}$\,cm$^{-2}$}
(optically thin \mbox{$J$\,=\,2--1} lines), also at $T_{\rm k}$\,$=$\,60\,K,
and similar $n$(H$_2$) range, we predict
population inversions (masers, \mbox{$T_{\rm ex}$\,$<$\,0}, with small amplification factors, i.e., small \mbox{$|\,\tau\,|$}). The shaded areas in 
\mbox{Figs.~\ref{fig:grid_Tex} and \ref{fig:grid_tau}} 
 mark the parameter space of these two particular cases.

\subsection{Anomalous HCN \mbox{$J$\,=\,1--0} HFS line ratios $R_{12}$ and $R_{02}$}

Figure~\ref{fig:grid_R} shows the integrated line intensity ratios 
$R_{12}$ and $R_{02}$ that result from our
 grid of static cloud models.  Everywhere \mbox{$\tau_{\rm hfs}$\,$\gtrsim$\,1}, models 
 including line overlap
 (continuous curves) show  different $R_{12}$ and $R_{02}$ values than models that do not (dotted curves). For HCN column densities below $\sim$10$^{13}$\,cm$^{-2}$, the line opacities are  low, and the HCN\,$J$\,=\,1-0 HFS line ratios are always $R_{12}$\,$\simeq$\,0.6 and $R_{02}$\,$\simeq$\,0.2. This is the usual  case of interstellar H$^{13}$CN and
HC$^{15}$N lines. As line opacities increase, $R_{12}$ takes smaller (anomalous) values produced by
the transfer of population to the  \mbox{$J$\,=\,1, $F$\,=\,2} level. Hence, we expect that the  HCN emission from GMCs will show anomalous  $R_{12}$\,$<$\,0.6 ratios, nearly independently of $T_{\rm k}$,  for \mbox{$N$(HCN)\,$\geq$\,10$^{13}$\,cm$^{-2}$}. 
On the other hand, $R_{02}$ reaches values 
slightly above (or below) the optically thin limit of 0.2
depending on  physical conditions and line widths.

\begin{figure}[h]
\centering  
\vspace{-0.3cm} 
\includegraphics[scale=0.31, angle=0]{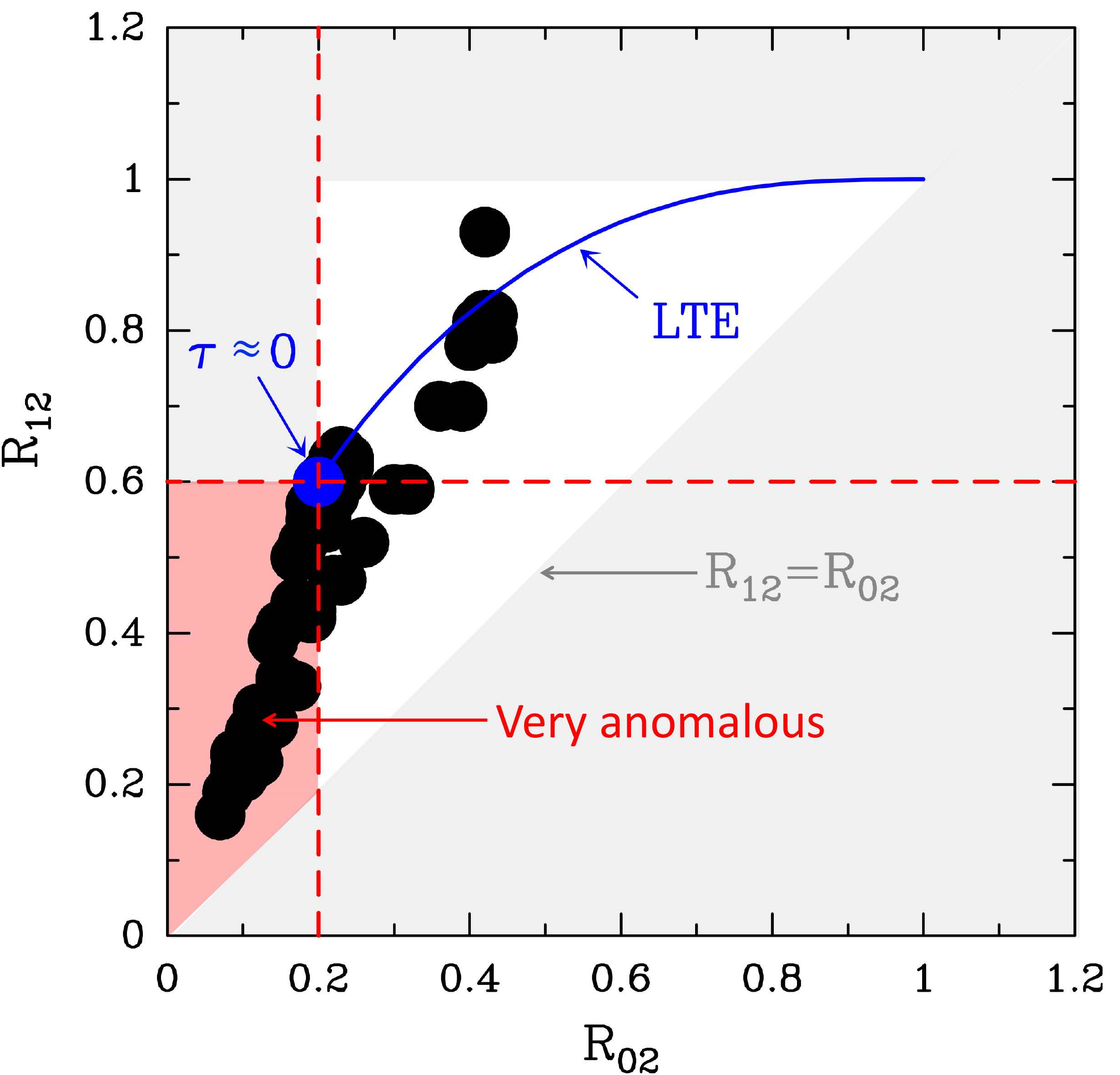}
\vspace{-0.2cm}
\caption{$R_{12}$ versus $R_{02}$ from our grid of 
 standard  models. The blue curve shows the expected LTE ratios as HCN line opacities
increase. The red shaded area shows
very anomalous   line intensity ratios
(\mbox{$R_{12}$\,$<$\,0.6} and \mbox{$R_{02}$\,$<$\,0.2}).
Gray shaded areas show regions of the \mbox{$R_{12}$--$R_{02}$} plane
that cannot be explained by single-component static cloud models.}
\label{fig:ratios_plane}
\end{figure}

Figure~\ref{fig:ratios_plane} summarizes the $R_{12}$ and $R_{02}$
values obtained from our grid of models. The blue curve shows the expected ratios
in LTE (single $T_{\rm ex}$) as HCN column densities increase. We note that the regions of the \mbox{$R_{12}$--$R_{02}$} plane: \mbox{[$R_{12}$\,$>$\,0.6 and $R_{02}$\,$<$\,0.2]},
\mbox{[$R_{12}$\,$<$\,$R_{02}$]}, 
and  \mbox{[$R_{12}$\,$>$\,1 and $R_{02}$\,$>$\,1]}
 can not be explained by these 
single-component static-cloud models. 
They can only be \mbox{interpreted} by the inclusion
of gas velocity gradients and \mbox{absorbing} envelopes of lower density gas
\citep[e.g.,][]{Gonzalez93,Magalhes:18}.

\begin{figure}[ht]
\centering   
\includegraphics[scale=0.71, angle=0]{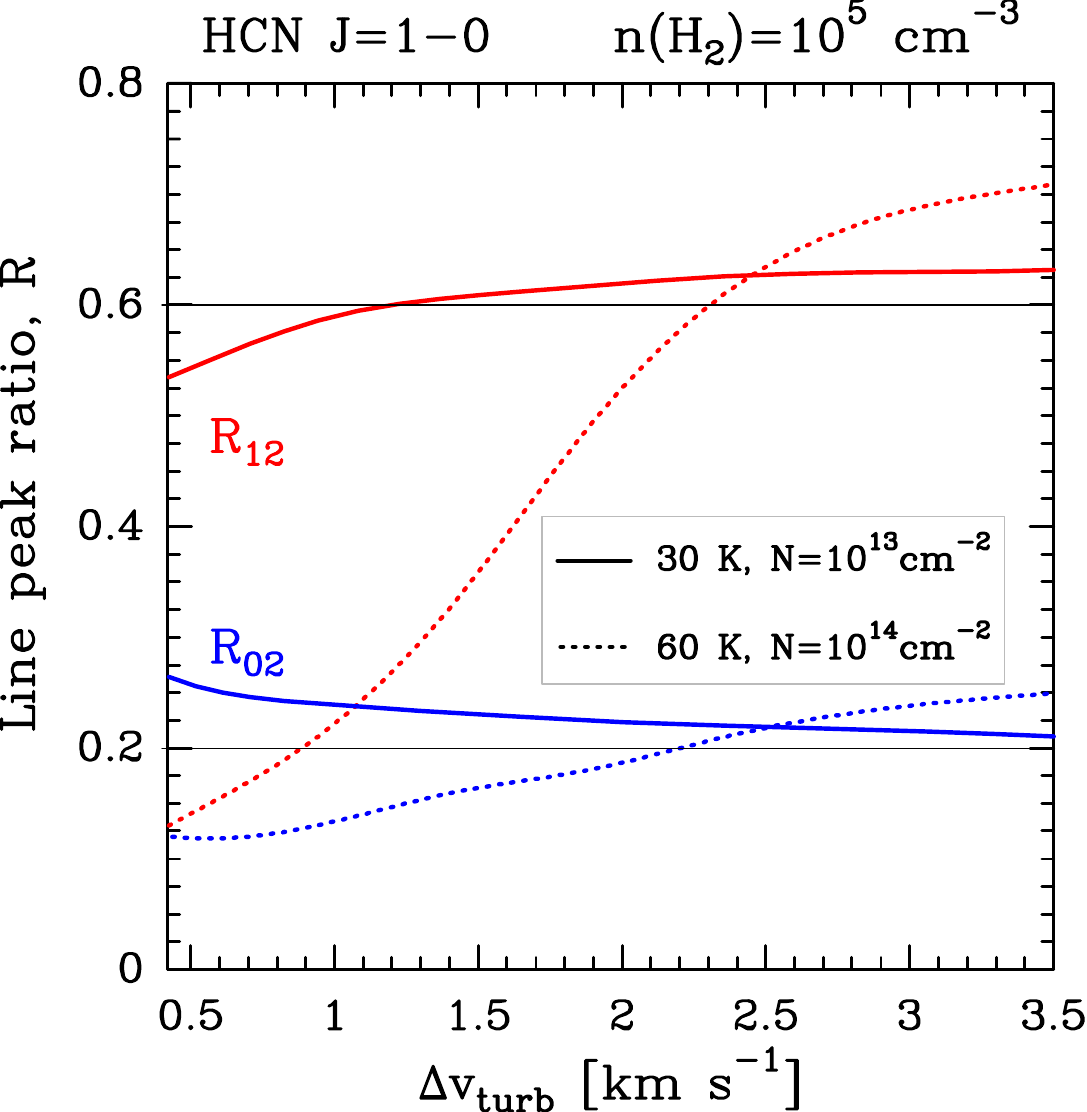}
\caption{Evolution of HCN $J$\,=\,1--0 HFS line peak ratios $R_{12}$ and $R_{02}$
for increasing microturbulent velocity dispersion (line widths).}
\label{fig:grid_n1e5_sigma_turb}
\end{figure}

Standard LVG models such as RADEX that do not treat line overlap
\mbox{\citep[e.g.,][]{Faure12}} are not capable of explaining these anomalous ratios or the
increased $T_{\rm ex,\,{\it F}=2-1}$ and reduced $T_{\rm ex,\,{\it F}=1-1}$
and $T_{\rm ex,\,{\it F}=0-1}$ values. 
At much higher HCN column densities ($>$10$^{16}$\,cm$^{-2}$) typical of the Orion hot core  \citep[e.g.,][]{Blake87,Schilke92}, all HFS lines become very opaque
and the intensity ratios tend to $R_{12}$\,$\rightarrow$\,1 and $R_{02}$\,$\rightarrow$\,1. As a corollary, we see that for very common physical conditions and moderate columns $N$(HCN),
the widely observed HCN \mbox{$J$\,=\,1--0} HFS lines show anomalous intensity ratios.
In this case, each HFS transition has a different  $T_{\rm ex,\,hfs}$ value, especially the strongest \mbox{$J$\,=\,1--0, $F$\,=\,2-1} line. This is a caution against the blind application
of automatic HFS  fitting programs that are precisely  based on the assumption of
optically thin 1:5:3 intensity ratios and the same $T_{\rm ex}$ for all \mbox{$J$\,=\,1--0} HFS components.

\subsection{Role of $\Delta v_{\rm turb}$ in the anomalous HFS line intensity ratios}

Figure~\ref{fig:grid_n1e5_sigma_turb} shows the effects of line overlap for increasing \mbox{microturbulent} velocity dispersion
(increasing line widths) in  dense gas models with  \mbox{$n$(H$_2$)\,=\,10$^{5}$\,cm$^{-3}$}. When intrinsic line widths increase, \mbox{$J$\,=\,1--0} HFS lines start to blend. Hence, this figure shows   $R_{12}$ and $R_{02}$
as line peak ratios (not as integrated line intensity ratios). 
In general, increasing
\mbox{$\Delta$$v_{\rm turb}$} reduces \mbox{$\tau_{\rm hfs}$}
and makes the intensity ratios less anomalous.
At low line opacities,  $R_{12}$ and $R_{02}$ do not depend much on \mbox{$\Delta$$v_{\rm turb}$} (continuous curves in Fig.~\ref{fig:grid_n1e5_sigma_turb}, models with  \mbox{$N$(HCN)\,=\,10$^{13}$\,cm$^{-2}$}, 
\mbox{$T_{\rm k}$\,=\,30\,K}, and \mbox{$\Delta$$v_{\rm turb}$\,$>$1\,km\,s$^{-1}$}).
Models with \mbox{$N$(HCN)\,=\,10$^{14}$\,cm$^{-2}$} and \mbox{$T_{\rm k}$\,=\,60\,K}, however, still show  anomalous ratios 
at \mbox{$\Delta$$v_{\rm turb}$\,$\gtrsim$\,2\,km\,s$^{-1}$} (dashed curves in Fig.~\ref{fig:grid_n1e5_sigma_turb}) because \mbox{$\tau_{F=2-1}$\,$>$\,1} and \mbox{$\tau_{F=1-1}$\,$>$\,1} at all \mbox{$\Delta$$v_{\rm turb}$}. 

\begin{figure}[ht]
\centering   
\includegraphics[scale=0.47, angle=0]{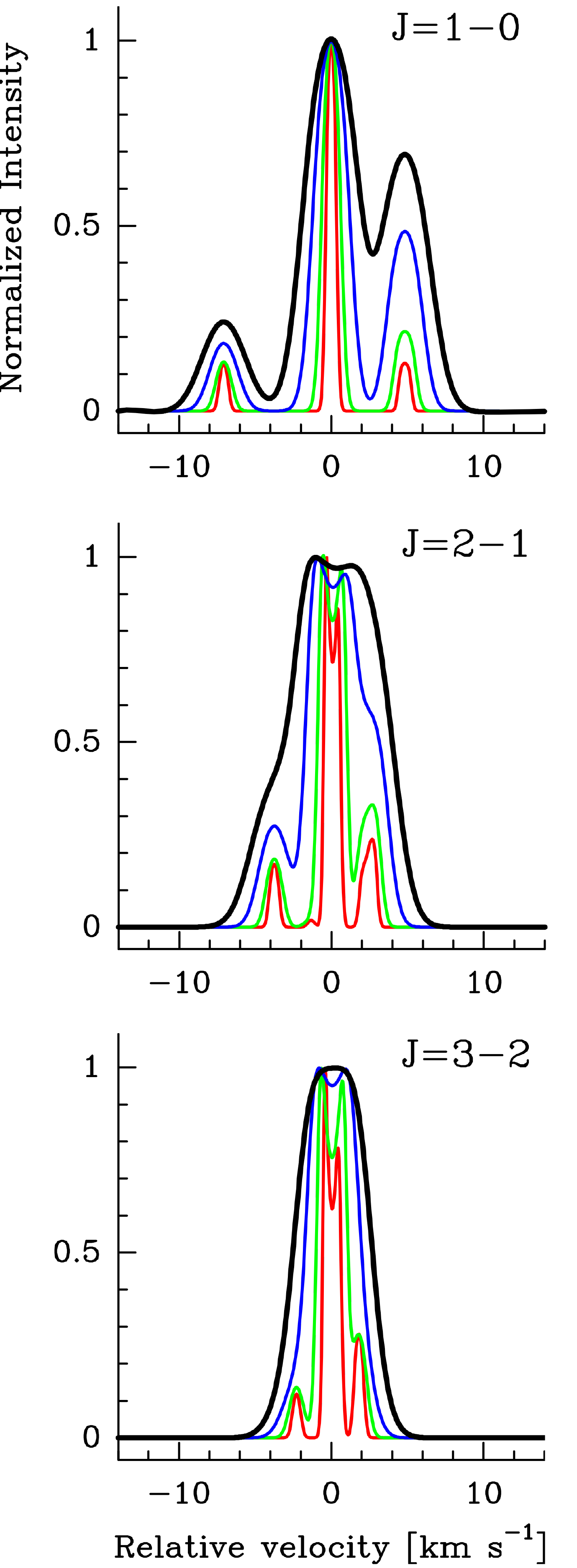}
\caption{Evolution of HCN  line profiles for increasing
turbulent velocity dispersion  with  fixed \mbox{$N$(HCN)\,=\,10$^{14}$\,cm$^{-2}$}, \mbox{$T_{\rm k}$\,=\,60\,K}, and \mbox{$n$(H$_2$)\,=\,10$^5$\,cm$^{-3}$}.
Microturbulent line widths are  \mbox{$\Delta$$v_{\rm turb}$\,$=$\,0.4\,km\,s$^{-1}$} (red curves),
\mbox{1\,km\,s$^{-1}$} (green), \mbox{2\,km\,s$^{-1}$} (blue), and \mbox{3\,km\,s$^{-1}$} (black).}
\label{fig:compa_spectra_HCN_grid_N1e14_sig}
\end{figure}

Figure~\ref{fig:compa_spectra_HCN_grid_N1e14_sig} shows the evolution of HCN $J$\,=\,1--0, 2--1, and 3--2 line profiles as the gas  velocity dispersion increases.  \mbox{$J$\,=\,1--0 HFS} line profiles clearly show the effects of line overlap in static and uniform clouds: brighter \mbox{$F$\,=\,2--1}, fainter \mbox{$F$\,=\,1--1} and, to a lesser extent,  fainter \mbox{$F$\,=\,0--1} lines.
In addition, HFS lines of higher $J$ transitions can also show anomalous ratios. 
Actually, observations of low-mass star-forming cores do show anomalous  \mbox{HCN $J$\,=\,3--2} HFS line intensity ratios \citep[e.g.,][]{Loughnane12}.
This rotational transition has six HFS lines, but  the four central ones are blended 
and cannot be resolved. This gives the impression of three lines
with relative intensity ratios of  \mbox{1:25:1} in the optically thin limit
(compare the \mbox{HCN $J$\,=\,3--2} line profiles in the lower panel of \mbox{Fig.~\ref{fig:lines_sketch}} with those in the right panel of \mbox{Fig.~\ref{fig:compa_spectra_HCN_grid_N1e14_sig}}).
 Higher HCN column densities and the inclusion of cloud velocity gradients will produce a greater variety of $R_{12}$ and $R_{02}$ values \citep[][]{Zinchenko87,Gonzalez93,Turner97,Mullins16,Magalhes:18}. This sensitivity to physical conditions means that HCN  can be a powerful probe  if their HFS lines are properly modeled.

\begin{figure*}[h]
\centering   
\includegraphics[scale=0.715, angle=0]{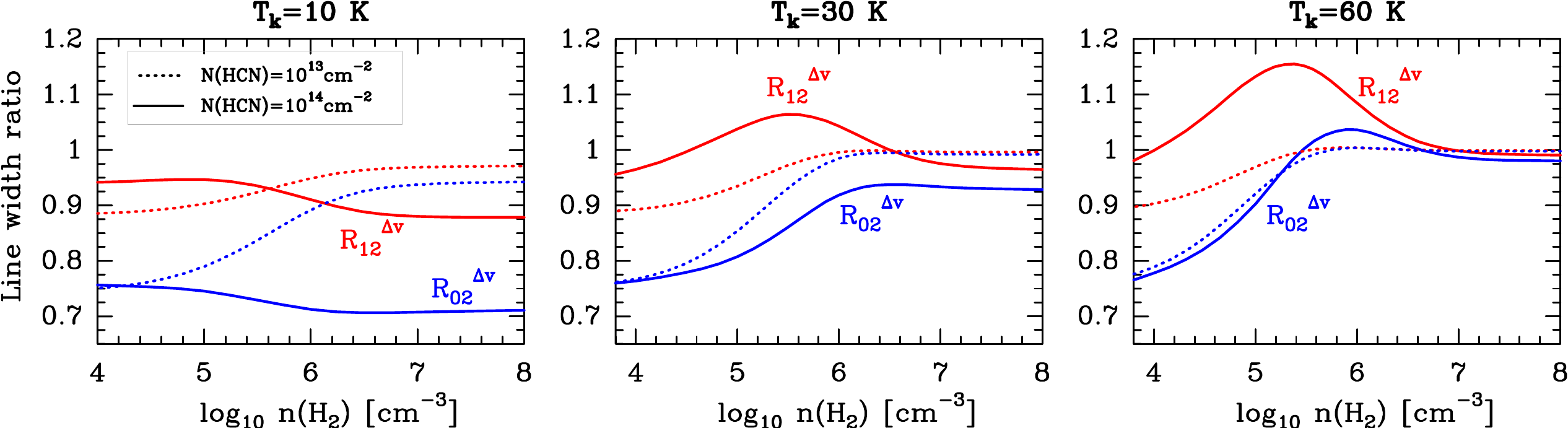}
\caption{HCN \mbox{$J$\,=\,1--0}  HFS line width ratios \mbox{$R_{02}^{\Delta v}$\,=\,$\Delta v_{F=0-1}$/$\Delta v_{F=2-1}$} and \mbox{$R_{12}^{\Delta v}$\,=\,$\Delta v_{F=1-1}$/$\Delta v_{F=2-1}$} for
models including line overlap.}
\label{fig:grid_dv}
\end{figure*}

\begin{figure*}[h]
\centering   
\includegraphics[scale=0.68, angle=0]{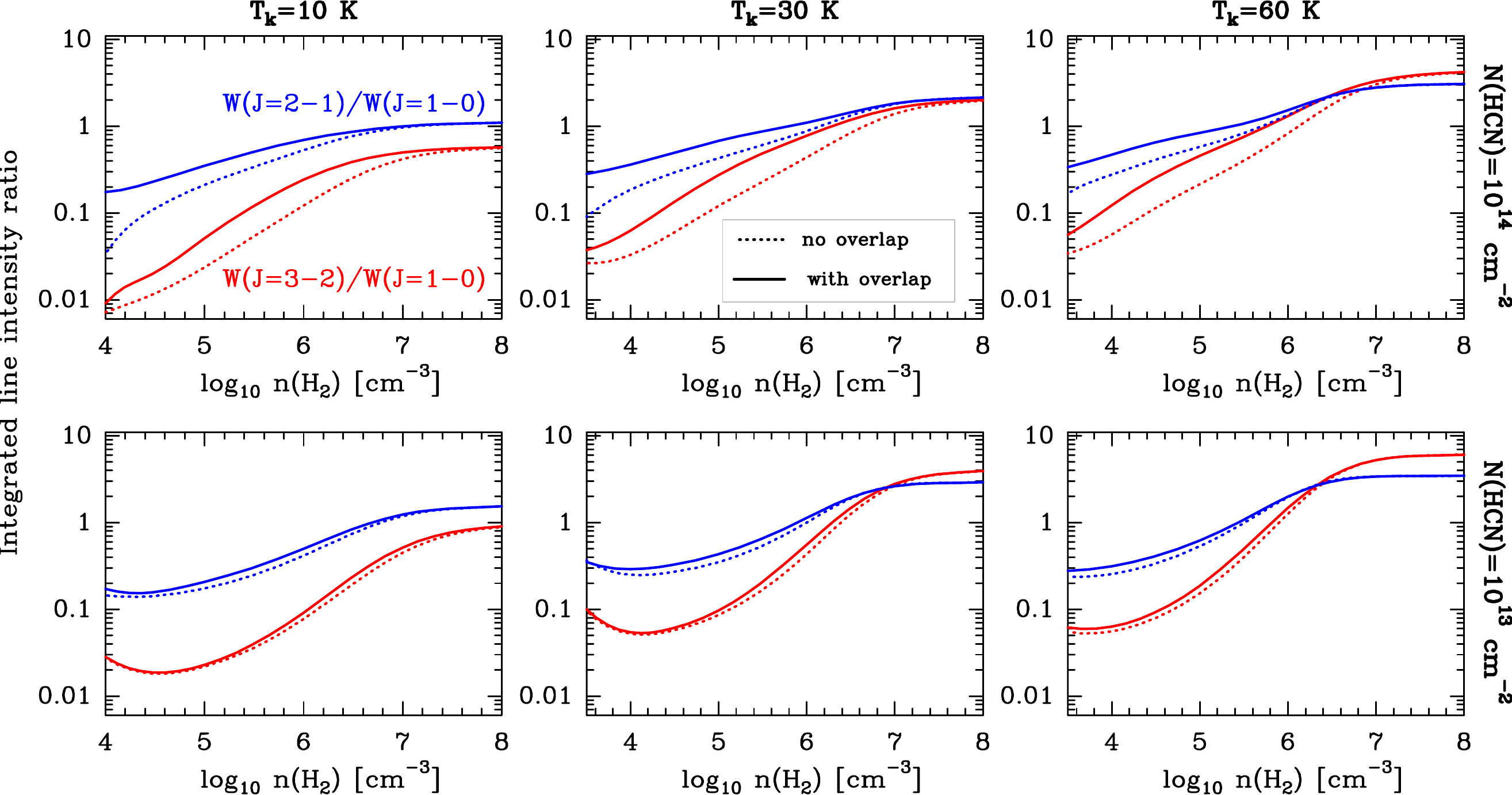}
\caption{Effect of HFS line overlap in the HCN rotational excitation. Each panel shows the integrated line intensity ratios (all HFS lines) \mbox{$W$($J$\,=\,2--1)/$W$($J$\,=\,1--0)} (blue curves) and \mbox{$W$($J$\,=\,3--2)/$W$($J$\,=\,1--0)} (red curves)  for
different $N$(HCN) and $T_{\rm k}$ values, 
and $W$ in units$^5$  of K\,km\,s$^{-1}$.}
\label{fig:grid_32_10}
\end{figure*}

\subsection{Line width anomalies $R_{12}^{\Delta v}$ and $R_{02}^{\Delta v}$}

In many instances not only the intensity ratios are anomalous,  also the HFS line opacity ratios become
anomalous  (Fig.~\ref{fig:grid_tau}). The most common combination
in our grid of static cloud models is 
  \mbox{$\tau_{F=2-1}$\,$>$\,$\tau_{F=1-1}$\,$>$\,$\tau_{F=0-1}$}. Therefore, as the HFS lines become optically thick,  opacity broadening will generally lead to 
\mbox{$\Delta v_{F=2-1}$\,$>$\,$\Delta v_{F=1-1}$\,$>$\,$\Delta v_{F=0-1}$} line widths.
For two optically thick lines with the same thermal and  microturbulent broadening, their line width ratio
only depends on their relative opacities \mbox{\citep[][]{Phillips79}}.
Hence, while in the optically thin limit  
\mbox{$R_{02}^{\Delta v}$\,=\,$\Delta v_{F=0-1}$/$\Delta v_{F=2-1}$\,=\,1} and
\mbox{$R_{12}^{\Delta v}$\,=\,$\Delta v_{F=1-1}$/$\Delta v_{F=2-1}$\,=\,1}, these 
line width ratios also change due to anomalous line opacities produced by line overlap effects. 
Indeed, \cite{Loughnane12} presented HCN $J$\,=\,1--0 HFS detections toward G333 massive cores 
showing anomalous HFS line width ratios. 
In our grid of models, we find \mbox{$R_{02}^{\Delta v}$} ranging from $\sim$0.71 to $\sim$1.03 and
 \mbox{$R_{12}^{\Delta v}$}  ranging from $\sim$0.87 to $\sim$1.2 \mbox{(Fig.~\ref{fig:grid_dv})}.
This implies that  the satellite line  \mbox{$F$\,=\,1--1} (and less frequently the other satellite line \mbox{$F$\,=\,0--1}) can  be broader than the main HFS line \mbox{$F$\,=\,2--1}. 
In particular, we predict  \mbox{$R_{12}^{\Delta v}$\,$>$\,1} in warm 
 (\mbox{$T_{\rm k}$\,$\geq$\,30\,K}) and dense  (\mbox{$>$\,10$^4$\,cm$^{-3}$}) 
 gas if 
 the HCN column density is large, \mbox{$N$(HCN)\,$\gtrsim$\,10$^{14}$\,cm$^{-2}$}. 
As a complementary corollary, we conclude that the assumption of uniform HCN line
widths in HFS  fitting programs may not always be justified. We recommend observers to check the 
line width of each \mbox{$J$\,=\,1--0} HFS line individually because they may carry information
about the HCN excitation conditions. When it varies, the true gas velocity dispersion should be extracted
from  optically thinner and not overlapped lines emitted by chemically related species.

\begin{figure*}[ht]
\centering   
\includegraphics[scale=0.69, angle=0]{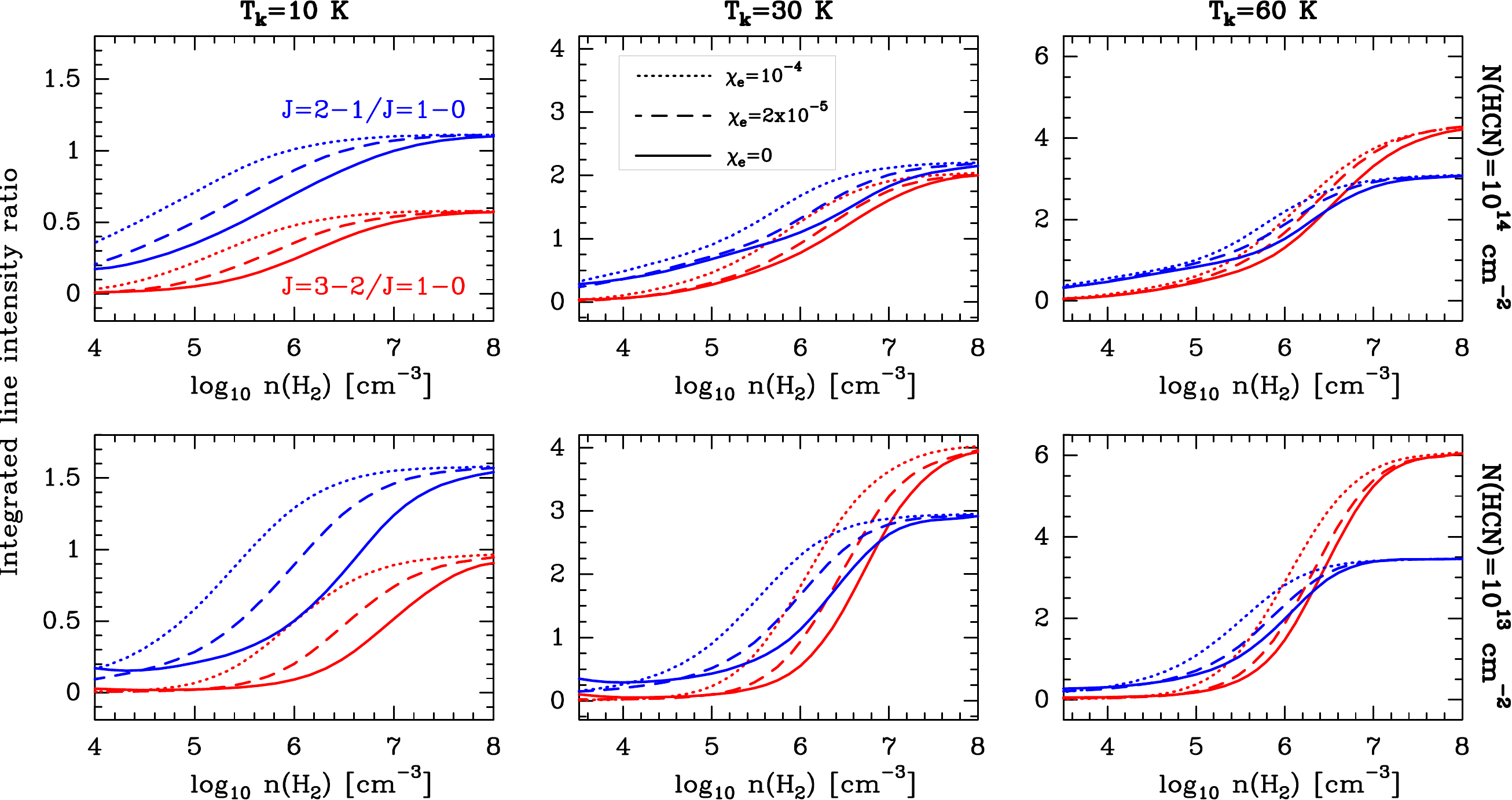} 
\caption{Effects of electron excitation in the HCN rotational excitation. 
Each panel shows the integrated line intensity ratios \mbox{W($J$\,=\,2--1)/W($J$\,=\,1--0)} (blue curves) and \mbox{W($J$\,=\,3--2)/W($J$\,=\,1--0)} (red curves)  
for different $N$(HCN) and $T_{\rm k}$  (with $W$ in units$^5$  of K\,km\,s$^{-1}$).}
\label{fig:grid_32_10_xe}
\end{figure*}

\subsection{Line overlaps and the excitation of higher-$J$  HCN lines}

A common misconception is that line overlap effects do not need to be treated if the 
cloud velocity dispersion  is such that the HFS lines are  \mbox{sufficiently} broad
 and, thus, the HFS structure is not spectrally  \mbox{resolved} by observations
(e.g., all HNC rotational lines and  rotationally excited HCN lines). 
Figure~\ref{fig:grid_32_10} shows the total  (all HFS lines) integrated intensity ratios   \mbox{$W$($J$\,=\,2--1)/$W$($J$\,=\,1--0)} (blue curves) and
\mbox{$W$($J$\,=\,3--2)/$W$($J$\,=\,1--0)} (red curves) extracted from our
 grid of  models (with $W$ in 
 units\footnote{To obtain the integrated line surface brightness ratios
 \mbox{$I$($J$\,=\,2--1)/$I$($J$\,=\,1--0)} and
 \mbox{$I$($J$\,=\,3--1)/$I$($J$\,=\,1--0)} with $I$ in units
 of \mbox{erg\,s$^{-1}$\,cm$^{-2}$\,sr$^{-1}$}, one has to 
 multiply the  ratios \mbox{$W$($J$\,=\,2--1)/$W$($J$\,=\,1--0)}
 and  \mbox{$W$($J$\,=\,3--2)/$W$($J$\,=\,1--0)}  by \mbox{$(\nu_{2-1}/\nu_{1-0})^3$$\simeq$\,8} and
 \mbox{$(\nu_{3-2}/\nu_{1-0})^3$$\simeq$\,27}.}
   of K\,km\,s$^{-1}$). When line overlap effects are relevant, the
 excitation temperatures
of the \mbox{$J$\,=\,2--1} and \mbox{$J$\,=\,3--2} HFS lines typically increase. This leads to 
intensity ratios  \mbox{$W$($J$\,=\,2--1)/$W$($J$\,=\,1--0}) and \mbox{$W$($J$\,=\,3--2)/$W$($J$\,=\,1--0)} 
that are higher, by a factor of about two in our range of standard physical conditions, than the intensity ratios computed ignoring line overlaps. Hence,
line overlap effects changes the $T_{\rm ex}$ of the \mbox{higher--$J$} lines. These differences will be more pronounced at higher line opacities.
We conclude that a precise analysis of the rotationally excited lines (and their intensity ratios) of  abundant species such as HCN, HNC, or N$_2$H$^+$ requires that  their HFS line overlaps are treated \mbox{\citep[][]{Daniel08,Keto10}}. If not,  parameters such as the gas density can  be overestimated.
A possible example is \mbox{OMC-1} clump in Orion\,A, where observations reveal relatively extended \mbox{HCN~$J$\,=\,6--5} \citep{Goicoechea19} and  \mbox{N$_2$H$^+$~$J$\,=\,7--6}  line emission \citep{Hacar20}.

\subsection{Role of electron excitation}\label{sec:role_elec}

In Sects.~\ref{sec:elec_col} and \ref{sec:crit_dens}, we anticipate that electron collisions
 play a  role in HCN excitation when  the gas ionization fraction
 is above the critical value of $\chi_{\rm cr}^{*}$($e^-$)\,$>$\,10$^{-5}$ \citep[see  also][]{Dickinson77,Liszt12,Goldsmith17}. Such high electron
 abundances are typical of the illuminated rims of GMCs  \mbox{\citep[their PDRs,][]{Goicoechea09,Cuadrado19}} and of more extreme GMCs in galaxy nuclei, typically irradiated by enhanced doses of cosmic- and X-rays  \citep[][]{Maloney96,Meijerink05}. In addition,  lower density and lower UV-illumination translucent clouds also
have high electron abundances \citep[][]{Black91,Hollenbach91}. 
In this section, we investigate the role of HFS-resolved \mbox{HCN--e$^-$} collisions 
in this kind of environments.

\begin{figure*}[ht]
\centering   
\includegraphics[scale=0.575, angle=0]{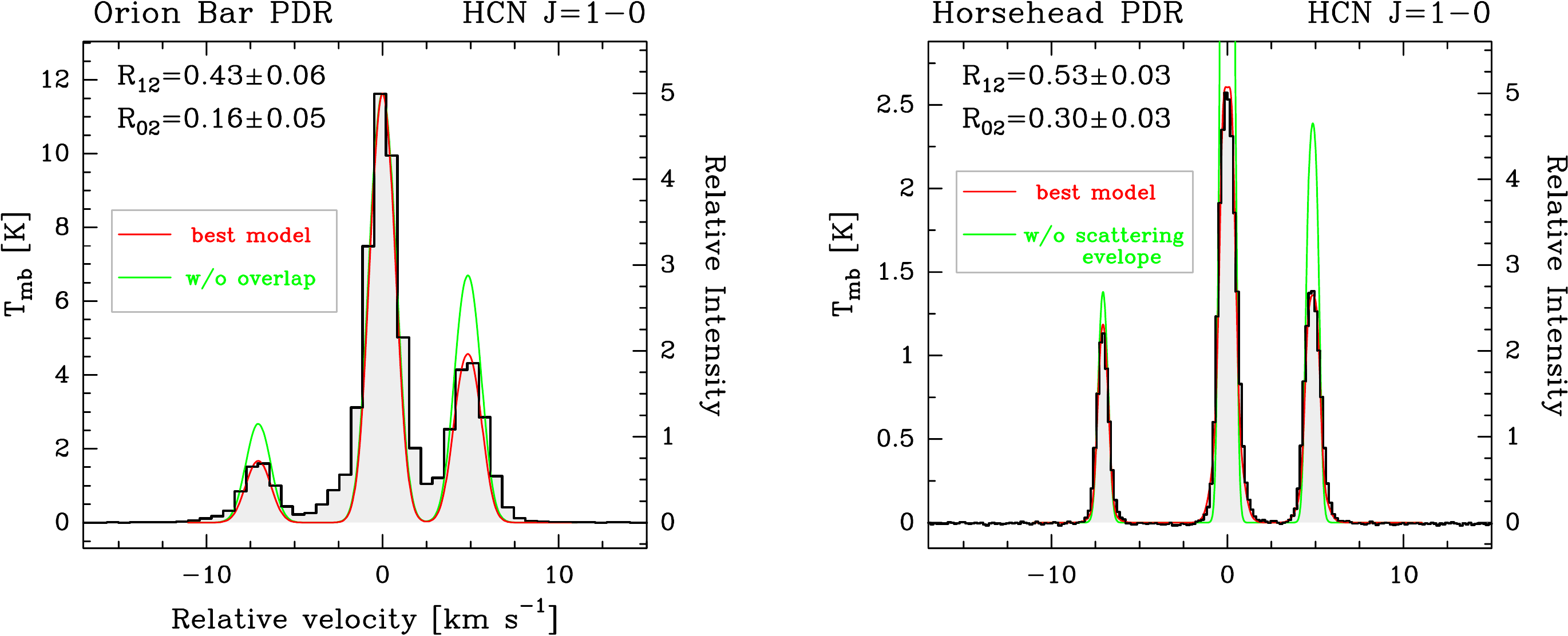}
\caption{IRAM\,30\,m observations and models of the \mbox{HCN\,$J$\,=\,1--0} HFS lines
toward the Orion Bar and the Horsehead PDRs. The intensity scale
in the right axes
is normalized to make clear that the observed line emission differs from 
the optically thin LTE line ratios \mbox{1:5:3} in both sources.} 
\label{fig:bar_hh}
\end{figure*}

Figure~\ref{fig:grid_R} shows the evolution of the intensity ratios
$R_{12}$ and $R_{02}$ in line overlap models with $\chi_{\rm e}$\,=\,10$^{-4}$ (dashed curves). These curves show about the maximum expected variation of $R_{12}$ and $R_{02}$. 
In other words,  lower \mbox{$\chi_{\rm e}$} 
abundances will result in less differences (ratios similar to the case without 
electron collisions). In the optically thin limit, the intensity ratios
 do not change much, and are \mbox{$R_{12}$\,$\simeq$\,0.6} and \mbox{$R_{02}$\,$\simeq$\,0.2}. 
As line opacity and overlap effects become important, 
electron collisions lead to intensity ratios $R_{12}$ that can be up to \mbox{$\sim$25\,$\%$} larger (in models with \mbox{$N$(HCN)\,=\,10$^{14}$\,cm$^{-2}$}; dashed curves) than when electron excitation does not play a role.
However, the  intensity ratio $R_{02}$ in these models is nearly independent of $\chi_{\rm e}$.
 Models with \mbox{$N$(HCN)\,=\,10$^{13}$\,cm$^{-2}$}
and low densities, \mbox{$n$(H$_2$)\,$\lesssim$\,10$^4$\,cm$^{-3}$}, 
result in marginally thick \mbox{$J$\,=\,1--0} HFS lines
and intensity ratios \mbox{$R_{12}$\,$>$\,0.6} and \mbox{$R_{02}$\,$>$\,0.2}.  Electron \mbox{excitation}  reduces their line opacities and, thus,
the intensity ratios $R_{12}$ and $R_{02}$ are lower
(closer to 0.6 and 0.2, respectively) than when electrons are not included (see also Sect.~\ref{sec:envelopes}).

We recall that excitation of polar molecules by electron collisions has
a strong dipole character (i.e., dominated by $|\,\Delta$$J\,|$\,=\,1 collisions) and therefore obeys different propensity rules than HCN
collisions with H$_2$; for which \mbox{$|\,\Delta$$J\,|$\,=\,2} collisions are relevant. This  explains the
less anomalous  $R_{12}$ values when electron collisions dominate.

As the gas density decreases
below  \mbox{$n_{\rm cr}$(H$_2$)}, electron  collisions compete with
H$_2$ collisions (at the lowest densities, electron collisions dominate). The major effect of electron collisions
is to produce more intense HCN \mbox{$J$\,=\,1--0} emission when 
\mbox{$n$(H$_2$)\,$<$\,$n_{\rm cr}$(H$_2$; $J$\,=\,1--0)} \citep[][]{Goldsmith17}. The intensity enhancement goes from factors of about two if 
\mbox{$\chi_{\rm e}$\,=\,2$\cdot$10$^{-5}$}, to factors of about ten if 
\mbox{$\chi_{\rm e}$\,=\,10$^{-4}$} (see Fig.~\ref{fig:grid_W_xe}). 
Figure~\ref{fig:grid_32_10_xe} shows that for the same H$_2$ density, electron collisions 
increase the population of the excited rotational levels, which enhances the 
\mbox{$W$($J$=2--1)/$W$($J$=1--0)} and \mbox{$W$($J$=3--2)/$W$($J$=1--0)} intensity ratios. 
The maximum effect is seen around \mbox{$n$(H$_2$)\,$\simeq$\,$n_{\rm cr}$(H$_2$; $J$\,=\,1--0)}, whereby H$_2$ gas densities
 are not too high to thermalize
the intensity ratios independently of $\chi_{\rm e}$, but not too low so that
the collisional rate is sufficiently high to appreciably populate the  excited levels $J$\,=\,2 and $J$\,=\,3. This means that detecting extended \mbox{HCN\,$J$\,=\,2--1}  emission
may not always imply the presence of very dense gas,
\mbox{$n_{\rm cr}$(H$_2$; $J$\,=\,2--1)\,$\simeq$ a few 10$^6$\,cm$^{-3}$},
 but lower density  gas with high ionization fractions.
Hence, a precise measurement of H$_2$ densities and
$\chi$(HCN) abundances  would greatly benefit from an estimation of the
 electron abundance of the gas where HCN emits \citep[e.g.,][]{Bron21}.

\section{Applications to the HCN emission from GMCs}\label{Sec-applications}

We conclude our study by applying our  models and new collisional rates
to specific warm gas environments in GMCs.

\subsection{Anomalous HCN emission from dense PDRs}

In this section, we model the \mbox{HCN\,$J$\,=\,1--0} HFS spectra of two prototypical  dense PDRs in  Orion\,A and B clouds: the edges of the Orion Bar and \mbox{Horsehead} nebula, both observed in spectral line surveys obtained with the IRAM\,30\,m telescope;
see \cite{Cuadrado_2015} and \cite{Pety_2012}, respectively.

The Orion Bar (strongly UV-irradiated\footnote{$G_0$ is the flux of far-UV photons (\mbox{$E$\,$<$\,13.6\,eV}) in
units of the Habing field. That is, \mbox{$G_0$\,$=$\,1} is equal to 1.6$\cdot$10$^{-3}$\,erg\,cm$^{-2}$\,s$^{-1}$.} PDR with \mbox{$G_0$\,$\gtrsim$\,10$^4$}): observed  \mbox{$J$\,=\,1--0} HFS line intensity ratios  are both anomalous: \mbox{$R_{12}$\,=\,0.43\,$\pm$\,0.06} and \mbox{$R_{02}$\,=\,0.16\,$\pm$\,0.05}. These  ratios can  be explained by line overlap effects. We obtain a satisfactory fit for the following
physical conditions: \mbox{$n_{\rm H}$\,$\simeq$\,2$\cdot$10$^5$\,cm$^{-3}$}, 
\mbox{$T_{\rm k}$\,$\simeq$\,100\,K},
\mbox{$\sigma_{\rm turb}$\,$\simeq$\,0.7\,km\,s$^{-1}$},
 \mbox{$\chi_{\rm e}$\,$\simeq$\,10$^{-4}$}, and  \mbox{$N$(HCN)\,$\simeq$\,10$^{14}$\,cm$^{-2}$} 
\mbox{(beam averaged)}. 
In this single-component model assuming extended emission (red curve in the left 
panel of  Fig.~\ref{fig:bar_hh}), the rotationally excited lines have higher opacities
(\mbox{$\tau_{J=3-2}$\,$\simeq$\,7} at the overlapping  HFS lines
\mbox{$F$\,=\,4--3, 3--2, and 2--1},
and \mbox{$\tau_{J=2-1}$\,$\simeq$\,4} at  
\mbox{$F$\,=\,3--2 and 2--1} lines) than the  \mbox{$J$\,=\,1--0} HFS lines
(\mbox{$\tau_{J=1-0}$\,$\simeq$\,1}). The excitation of these lines
is subthermal, with \mbox{$T_{\rm ex,\,hfs}$\,$\simeq$\,10--20\,K\,$\leq$\,$T_{\rm k}$\,=\,100\,K}.
The green curve in the left panel of \mbox{Fig.~\ref{fig:bar_hh}} shows a model with the same input parameters but neglecting line overlap. This model produces different line intensity ratios $R_{12}$ and $R_{02}$. 

\textit{The Horsehead} (mildly illuminated PDR with \mbox{$G_0$\,$\simeq$\,100}): observed intensity ratios are anomalous, 
\mbox{$R_{12}$\,=\,0.53\,$\pm$\,0.03} and \mbox{$R_{02}$\,=\,0.30\,$\pm$\,0.03}.
We checked that given the warm gas  temperatures and moderate densities previously inferred
in this PDR: \mbox{$T_{\rm k}$\,$\approx$\,60--100\,K}
and \mbox{$n_{\rm H}$\,$\approx$\,2$\cdot$10$^4$--10$^5$\,cm$^{-3}$}
\citep[e.g.,][]{Guzman11,Pabst17}, a single-component cloud model can not \mbox{explain} these ratios. In addition, the  intensity of the \mbox{$F$\,=\,0--1} line relative to the other HFS lines, is too strong and its line width
(\mbox{$\Delta$$v_{F=0-1}$\,=\,0.8\,$\pm$\,0.1\,km\,s$^{-1}$}) 
is narrower than
those of the \mbox{$F$\,=\,2--1} and \mbox{$F$\,=\,1--1} lines
(\mbox{$\Delta$$v_{F=2-1}$\,$\simeq$\,$\Delta$$v_{F=1-1}$\,=\,1.1\,$\pm$\,0.1\,km\,s$^{-1}$}). These HCN line widths exceed those of the \mbox{H$^{13}$CN $J$\,=\,1--0} HFS lines (not shown; \mbox{$\Delta$$v_{\rm H^{13}CN}$\,=\,0.70\,$\pm$\,0.02\,km\,s$^{-1}$}). 
Hence, the \mbox{$F$\,=\,0--1/$F$\,=\,2--1} line width ratio is anomalous too, with
\mbox{$R_{02}^{\Delta v}$\,=\,0.7\,$\pm$\,0.1} and \mbox{$R_{12}^{\Delta v}$\,=\,1.0\,$\pm$\,0.1}.
These signatures suggest  optically thick  
\mbox{HCN $J$\,=\,1--0} lines  and  self-absorption of the
\mbox{$F$\,=\,2--1} and \mbox{$F$\,=\,1--1} lines. Thus, this is a more  complicated radiative transfer problem. A very likely scenario is that the edge of the Horsehead has a very steep density gradient, 
from  diffuse to dense gas, and/or that  line photons arising from the dense PDR are self-absorbed and scattered by a low density  envelope.

We  reproduce the anomalous \mbox{HCN $J$\,=\,1--0} HFS spectrum with a two-component model:
 a moderately  dense PDR  with \mbox{$N$(HCN)\,$\simeq$\,3$\cdot$10$^{13}$\,cm$^{-2}$}, \mbox{$n_{\rm H}$\,$\simeq$\,3$\cdot$10$^4$\,cm$^{-3}$}, 
\mbox{$T_{\rm k}$\,$\simeq$\,60\,K},
\mbox{$\sigma_{\rm turb}$\,$\simeq$\,0.2\,km\,s$^{-1}$}, and $\chi_{\rm e}$$\simeq$10$^{-4}$, 
\mbox{surrounded} by a lower excitation envelope with:
\mbox{$N$(HCN)\,$\simeq$\,1.5$\cdot$10$^{13}$\,cm$^{-2}$},
\mbox{$n_{\rm H}$\,$\simeq$\,4$\cdot$10$^3$\,cm$^{-3}$}, 
\mbox{$T_{\rm k}$\,$\simeq$\,30\,K}, and \mbox{$\chi_{\rm e}$\,$\lesssim$\,10$^{-5}$} 
\citep[typical of the UV--illuminated extended gas
in Orion\,B; e.g.,][]{Pety17}. 
The main effect of the envelope is to absorb a fraction of the optically thick \mbox{$F$\,=\,2--1} and \mbox{$F$\,=\,1--1} line photons emitted from the PDR and  to 
scatter them over large spatial scales. We note that
several studies have previously argued that resonant scattering
by low density envelopes or foreground clouds could explain the
  spatial distribution of  the 
subthermally excited and  optically thick emission   from abundant high dipole moment molecules such as HCO$^+$, HCN, or CS,
as well as their anomalous line intensities
\citep[sometimes comparable to those of their isototopologues;][]{Langer78,Walmsley82,Cernicharo84,Gonzalez93,Zinchenko93}.

In our model of the Horsehead, the total line opacities are 
\mbox{$\tau_{F=0-1}$\,$\simeq$\,1}, \mbox{$\tau_{F=2-1}$\,$\simeq$\,5}, and \mbox{$\tau_{F=2-1}$\,$\simeq$\,4}.
Since the observed HFS lines do not show  self-absorption dips,  the gas velocity dispersion in the
absorbing envelope needs to be larger (\mbox{$\sigma_{\rm turb}$\,$\simeq$\,0.5\,km\,s$^{-1}$}) than in the denser PDR.
The red curve in the right panel of \mbox{Fig.~\ref{fig:bar_hh}} shows the resulting  line profiles. In this model the excitation temperatures of the \mbox{$J$\,=\,1--0} HFS lines
are \mbox{$T_{\rm ex,\,hfs}$\,$\simeq$\,7--10\,K} in the PDR, and
\mbox{$T_{\rm ex,\,hfs}$\,$\simeq$\,3--4\,K} in the scattering envelope. The fact that 
$T_{\rm ex,\,hfs}$ in the low density envelope (very weak collisional excitation) 
is  higher than 3\,K is a pure scattering 
effect:  absorption and remission of line photons coming from the denser component
\citep[for cold dark cloud models, see][]{Gonzalez93}.
The green curve shows results of a model with the same input parameters
for the dense PDR, but no scattering envelope,
which results in very different intensity and line width ratios.

We finally note that the gas densities we infer in these PDRs could be slightly lower if the
HCN  emission arises from lower ionization fraction gas.
In forthcoming papers, we will discuss more realistic models (e.g., with gradients) and analyze  \mbox{multiple-$J$} HCN observations of Orion\,B (Santa-Maria et al. in prep.)
and of the Orion Bar PDR (Goicoechea et al. in prep.).

\subsection{HCN emission from the extended environment of GMCs and whether it contributes to the extragalactic star-formation rate relation}\label{sec:envelopes}

The physical conditions in small translucent  clouds: $n$(H$_2$) up to several 10$^3$~cm$^{-3}$,
\mbox{$T_{\rm k}$\,$\simeq$\,15--60\,K}, and \mbox{$\chi_{\rm e}$\,$\gtrsim$\,10$^{-5}$}
\citep[e.g.,][]{vD89,Snow06} resemble those found in the extended environment of GMCs. These regions are very important when
we consider the integrated emission from spatially unresolved GMCs in distant 
star-forming galaxies.
Although HCN line intensities at any specific position of the extended cloud environment
 would be much fainter than at the dense  
star-forming cores (\mbox{$n$(H$_2$)\,$\gtrsim$\,10$^{5}$\,cm$^{-3}$})  
\mbox{ -- where} $\chi$(e$^-$)  is usually not high enough
to collisionally excite HCN lines \mbox{\citep[e.g.,][]{Salas21} --} the much larger area of the extended cloud emission, the cloud envelope,  means that emission lines integrated over the entire GMC can be dominated by the lower-density extended cloud and not by the dense cores \mbox{\citep[e.g.,][]{Evans20,Miriam21}}. This widespread (tens of pc) and more translucent GMC  environment is typically illuminated by modest stellar UV fields,  \mbox{$G_0$\,$\simeq$\,2--100} \citep[e.g.,][]{Pineda13,Abdullah20} that are less extreme than the incident UV field in the dense star-forming clumps
 ($\sim$1\,pc scales) close
to young massive stars \citep[up to \mbox{$G_0$\,$\simeq$\,10$^5$};  e.g.,][]{Goicoechea19,Pabst21}.

\begin{figure}[t]
\centering   
\includegraphics[scale=0.85, angle=0]{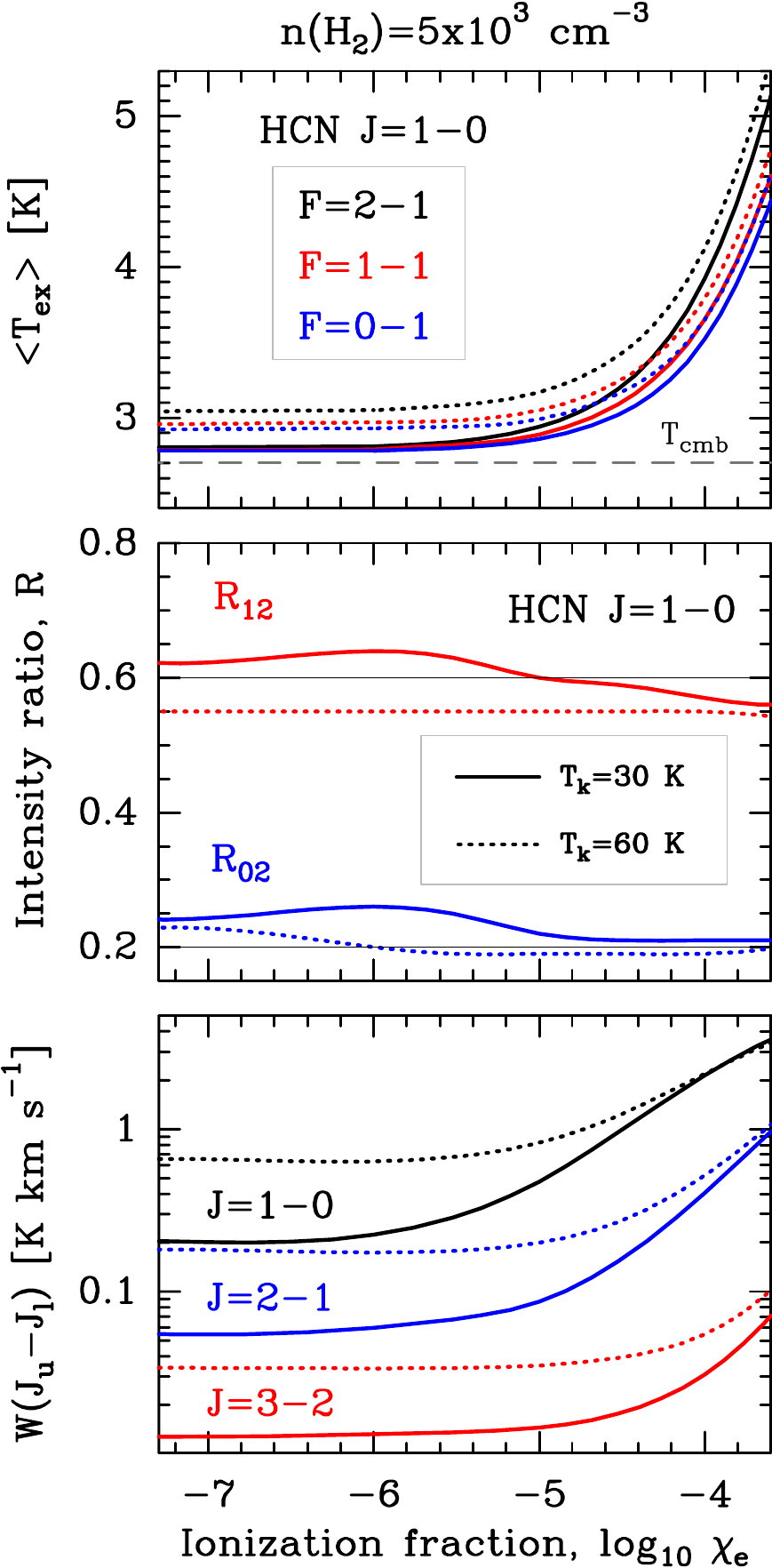}
\caption{Models of the extended and more translucent environment of GMCs
as a function of the ionization fraction. For $N$(HCN)\,=\,10$^{13}$\,cm$^{-2}$,  and $n$(H$_2$)\,=\,5$\cdot$10$^3$\,cm$^{-3}$ we show: \mbox{mean $T_{\rm ex}$(\mbox{$J$\,=\,1--0, $F_{\rm u}$-$F_{\rm l}$}) in the upper panel}; line intensity ratios $R_{02}$ and $R_{12}$ in the middle panel;  and total integrated line intensities \mbox{HCN $J$\,=\,1--0}, \mbox{2--1},  \mbox{3--2 in the lower panel. } Continuous and dotted curves show model results for \mbox{$T_{\rm k}$\,=\,30\,K} and
60\,K, respectively.} 
\label{fig:grid_translucent}
\end{figure}

Figure~\ref{fig:grid_translucent} shows model results appropriate to this 
GMC environment:
\mbox{$n$(H$_2$)\,$=$\,5$\cdot$10$^{3}$\,cm$^{-3}$}, \mbox{$\Delta v_{\rm turb}$\,=\,1\,km\,s$^{-1}$}, 
 \mbox{$T_{\rm k}$\,=\,30\,} and 60\,K,  and
$N$(HCN)\,=\,10$^{13}$\,cm$^{-2}$ \citep[the typical  HCN column density observed in translucent clouds;][]{Turner97,Godard10}.
Neglecting electron collisions results in very subthermal HCN emission: 
\mbox{$T_{\rm ex,\,hfs}$\,$\lesssim$\,3\,K} (see upper panel of \mbox{Fig.~\ref{fig:grid_translucent}}), with  emission levels \mbox{$I$($J$\,=\,1--0)} of  \mbox{$\sim$\,0.2\,K\,km\,s$^{-1}$} and 
\mbox{$\sim$\,0.6\,K\,km\,s$^{-1}$} for
\mbox{$T_{\rm k}$\,=\,30\,K} and 60\,K, respectively 
(lower panel of \mbox{Fig.\ref{fig:grid_translucent}}). For this choice of physical conditions, the opacity
of the main \mbox{$J$\,=\,1--0, $F$\,=\,2--1 HFS} line is \mbox{$\tau_{F=2-1}$\,$\simeq$\,2}. That is, line overlap
effects start to  matter. 

As the electron abundance increases, so does the excitation of the \mbox{$J$\,$\geq$\,1} levels.
For \mbox{$\chi_{\rm e}$\,$\simeq$\,10$^{-5}$},  the excitation temperature  of the \mbox{$J$\,=\,1--0} HFS lines
starts to rise despite the H$_2$ density being considerably lower than 
\mbox{$n_{\rm cr}$(H$_2$; $J$=1--0)}.
This particular $\chi_{\rm e}$ value leads
to a $I$($J$\,=\,1--0) enhancement  by a factor of two. 
 For \mbox{$\chi_{\rm e}$\,$\simeq$\,10$^{-4}$},   excitation temperatures rise to 
 \mbox{$T_{\rm ex,\,hfs}$\,$\simeq$\,4\,K}  and the line emission level increases up to
 a factor of ten, \mbox{$I$($J$\,=\,1--0)\,$\simeq$\,2\,K\,km\,s$^{-1}$}, compared to models that do not include electron excitation. Because levels $J$\,=2 and 3 are now more populated,
the opacity of the main \mbox{$J$\,=\,1--0, $F$\,=\,2--1 HFS} line decreases, but it is still \mbox{$\tau_{F=2-1}$\,$\simeq$\,1}.
We predict that  in this lower density and more translucent GMC environment, $R_{12}$  will be anomalous, \mbox{$R_{12}$\,$\lesssim$\,0.6}, and the  
 $R_{02}$ intensity ratio could be slightly above or below 0.2, again depending on
 the given physical conditions   
 and $\Delta v$ (\mbox{middle} panel of \mbox{Fig.\ref{fig:grid_translucent}}). Despite
 the low H$_2$ density, the \mbox{HCN $J$\,=\,2--1} line
 could be detectable, with  \mbox{$I$($J$\,=\,2--1)\,$\geq$\,0.1\,K\,km\,s$^{-1}$}
 (Fig.~\ref{fig:grid_W_xe}).
Therefore, electron collision are  very important at low H$_2$ densities, provided that HCN exists in gas with $\chi_{\rm e}$\,$>$\,10$^{-5}$, and they may contribute to 
the extended HCN emission in GMCs
 \citep[see also][]{Goldsmith17}. 
 
 By adopting as our references $\chi_{\rm e}$\,$=$\,2$\cdot$10$^{-5}$ and \mbox{$T_{\rm k}$\,=\,30\,K}, we
predict HCN line emission levels of  \mbox{$I$($J$\,=\,1--0)\,$\simeq$\,0.7\,K\,km\,s$^{-1}$}. For a spherical cloud of 10\,pc (or 100\,pc) diameter and uniform 
  emission, these numbers  imply a 
  integrated HCN $J$\,=\,1--0 line luminosity 
\citep[in units of \mbox{K\,km\,s$^{-1}$\,pc$^{2}$};][]{Gao2} of 
\mbox{$L_{\rm HCN}$\,$\approx$\,55\,K\,km\,s$^{-1}$\,pc$^{2}$}  
(\mbox{or $\approx$\,5500\,K\,km\,s$^{-1}$\,pc$^{2}$}) considering only this extended and translucent
\mbox{HCN emission}.

HCN $J$\,=\,1--0 observations of a large sample of star-forming galaxies
(from normal spirals to more  extreme ultraluminous \mbox{infrared} galaxies) find the following tight relationship:
\begin{equation}
\label{eq:sfr}
\dot{M}_{SFR}\,\approx\,1.8\cdot10^{-7}  (L_{\rm HCN}/ {\rm K\,km\,s^{-1}\,pc^{2}})\,\, 
{\rm[M_{\odot}\,yr^{-1}]},
\end{equation}
where $\dot{M}_{SFR}$ is the  star-formation rate (SFR) 
and $L_{\rm HCN}$ is (\mbox{assumed} to be) emitted by the dense molecular gas reservoir  \mbox{\citep{Gao04}}.
Inserting our $L_{\rm HCN}$ value in Eq.~\ref{eq:sfr}, we obtain
\mbox{$\dot{M}_{SFR}$\,$\approx$\,10$^{-5}$\,$M_{\odot}$\,yr$^{-1}$} for a 10\,pc cloud
(\mbox{$\approx$\,10$^{-3}$\,$M_{\odot}$\,yr$^{-1}$} for a 100\,pc cloud). 
We recall that these SFRs simply assume that $L_{\rm HCN}$
mostly arises from our toy model
 low-surface-brightness GMC environment.
Still, the resulting $\dot{M}_{SFR}$ rates are of the order of those inferred,
from other observational tracers,  toward galactic GMCs \mbox{\citep[e.g.,][]{Lada10}}.

This similitude would imply that for some galaxies, 
 the extragalactic  \mbox{$\dot{M}_{SFR}$--$L_{\rm HCN}$} correlation is
 not always dominated by emission  from dense molecular gas, 
 \mbox{$n_{\rm H}$\,$\gtrsim$\,10$^5$\,cm$^{-3}$}. This can be the case of  normal spiral galaxies
 in which $L_{\rm HCN}$  linearly correlates with $L_{\rm CO}$ (which is certainly dominated
by extended low-density molecular  gas) and also  with $L_{\rm FIR}$, a proxy of 
the SFR \mbox{\citep[][]{Gao04}}. Normal galaxies have low luminosity ratios 
 \mbox{$L_{\rm HCN}$/$L_{\rm CO}$\,$\simeq$\,0.02--0.06} that are interpreted as low fractions of dense molecular gas. These  \mbox{$L_{\rm HCN}$/$L_{\rm CO}$} values resemble the observed ratios in modest (low SFR)
 GMCs   such as Orion\,B  when  square-degree areas of the sky 
 are averaged \citep[\mbox{$L_{\rm HCN}$/$L_{\rm CO}$\,$\simeq$\,0.025}
 in \mbox{$\sim$40\,pc$^2$};][]{Pety17}.
 Indeed, disk GMCs such as Orion\,A and B show more spatially extended emission
 in HCN   than in other tracers, such as N$_2$H$^+$,
of  cold and dense gas  \citep[e.g.,][]{Pety17,Kauffmann17,Melnick_2020}.

 \mbox{Luminous} and ultraluminous infrared galaxies, however,  show an excess of HCN emission compared to CO  (\mbox{$L_{\rm HCN}$/$L_{\rm CO}$\,$>$\,0.06}), and
 only $L_{\rm HCN}$, not $L_{\rm CO}$, is closely correlated with $L_{\rm FIR}$ 
  \citep[][]{Gao04}. Hence, the HCN emission from these more extreme (very high SFR) galaxies  very likely traces a higher fraction of dense star-forming gas.
  According to our models,  a significant fraction of
the HCN luminosity at GMC scales may arise from the lower density extended component (the envelopes) of these clouds, at least in normal spirals.
Since $L_{\rm HCN}$ and $L_{\rm FIR}$ (SFR) tightly correlate over 
three orders of magnitude in galaxies, this scenario would 
 imply that the mass of the extended component scales with the mass of the dense star-forming cores. Careful analysis of on-going wide field  molecular emission 
 \mbox{surveys} of galactic GMCs (covering increasingly larger areas) are clearly needed to settle down this issue.

\begin{figure}[t]
\centering   
\vspace{-0.2cm} 
\includegraphics[scale=0.59, angle=0]{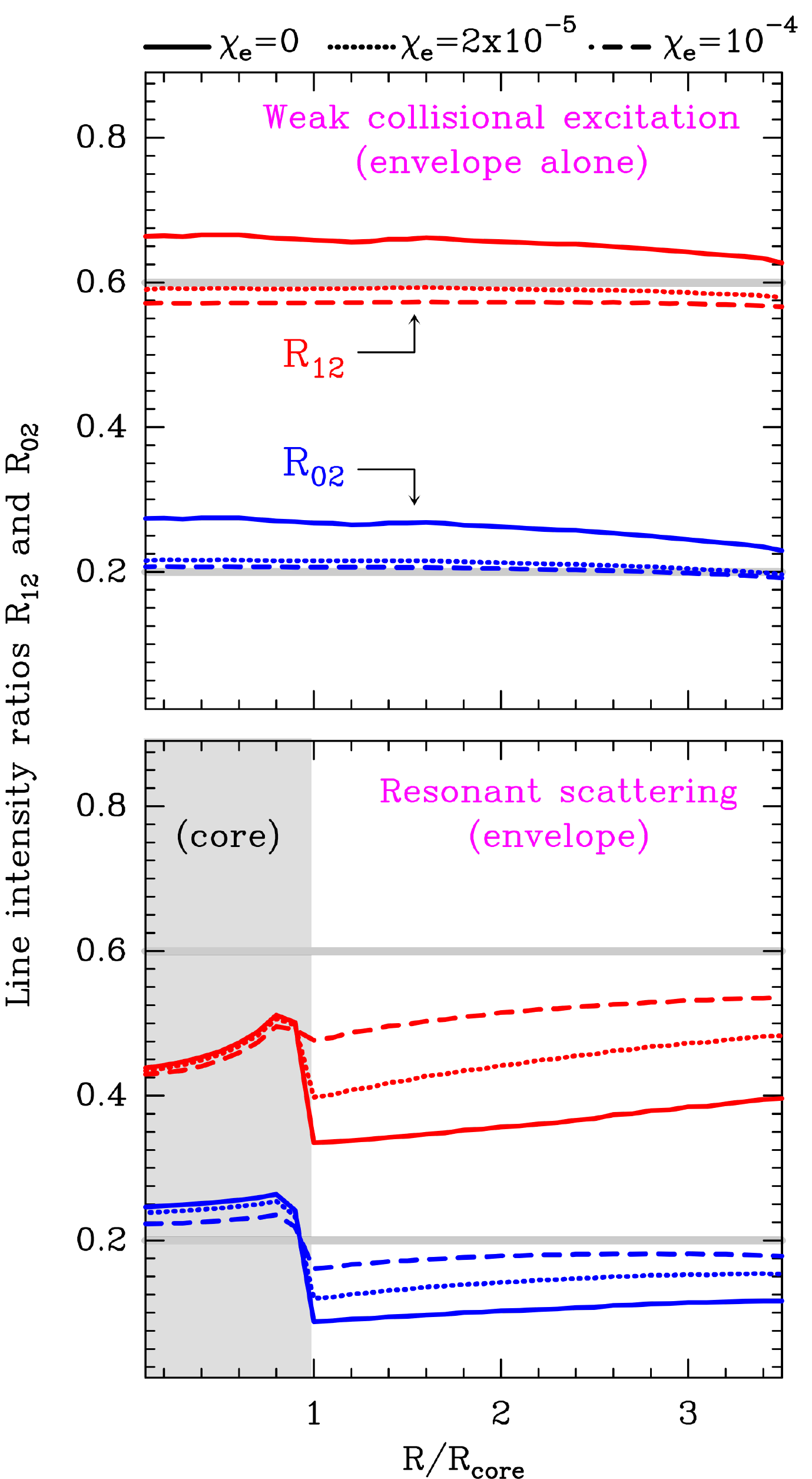}
\caption{HCN\,$J$\,=\,1--0 HFS intensity ratios $R_{12}$ (red curves)
and $R_{02}$ (blue curves) versus impact parameter  for two kind of extended
cloud environments (both with $n$(H$_2$)\,=\,5$\times$10$^3$\,cm$^{-3}$ and 
$T_{\rm k}$\,=\,30\,K). \mbox{\textit{Upper panel:}} Extended envelope alone, 
with $N$(HCN)\,=\,10$^{13}$\,cm$^{-2}$ along its diameter,
resulting
in (weakly) collisionally excited HCN emission. 
\textit{Lower panel:} Same envelope
surrounding a dense core, with $n$(H$_2$)\,=\,10$^5$\,cm$^{-3}$, and scattering
the bright \mbox{HCN\,$J$\,=\,1--0}  line emission arising from the core.} 
\label{fig:scattering} 
\end{figure}

Interestingly, the HCN\,$J$\,=1--0 HFS line intensity ratios
 $R_{12}$ and $R_{02}$ observed in local GMCs can 
 be used to quantify the amount of extended \mbox{HCN\,$J$\,=1--0 emission} that  arises from electron-assisted (weakly) collisionally excited low-density  gas,  versus  emission from high density cores resonantly scattered  over larger spatial scales by the low density cloud. In the latter case, the observed HCN luminosities will still reflect the fraction of dense molecular gas.
  Figure~\ref{fig:scattering} shows a model example of these two scenarios.
 The lower panel shows the predicted  intensity ratios $R_{12}$ 
and $R_{02}$  versus impact parameter (i.e., after ray tracing the spherical cloud) for a model of a dense core, with $n$(H$_2$)\,=10$^5$\,cm$^{-3}$ and 
$T_{\rm k}$\,=\,30\,K, surrounded by a lower density envelope,
with $n$(H$_2$)\,=\,5$\cdot$10$^3$\,cm$^{-3}$, four times larger than the core
(in these models the HCN abundance is fixed to \mbox{$\chi$(HCN)\,=\,3$\cdot$10$^{-9}$}). 
In this case, the bright \mbox{HCN\,$J$\,=1--0} emission arising from the core is scattered
by the low density envelope. For an impact parameter that crosses the envelope and not the core (i.e., an independent observation of the extended cloud) resonant scattering produces very anomalous line intensity ratios, with 
\mbox{$R_{12}$\,$\lesssim$\,0.5} and \mbox{$R_{02}$\,$<$\,0.2}. However, these ratios
appear to be less frequently observed in GMCs, at least on the spatial scales 
of previous observations \citep[e.g.,][]{Gottlieb75,Loughnane12}.
On the other hand, if the HCN emission intrinsically arises from low density gas far from
star forming cores, the weak collisional excitation drives the intensity ratios to
\mbox{$R_{12}$\,$\simeq$\,0.6}
and \mbox{$R_{02}$\,$\gtrsim$\,0.2} \mbox{(upper panel of Fig.~\ref{fig:scattering})}.
The presence of high electron abundances in the envelope does not change these conclusions
(dotted and dashed curves); however, as stated previously, it raises the \mbox{HCN\,$J$\,=1--0} intensities to detectable levels.

\section{Summary and conclusions}

We revisited the excitation  of HCN hyperfine lines 
considering \mbox{radiative} effects and line photon \mbox{exchanges} induced by HFS
line overlaps.
Accurate models of the HCN emission  require knowledge of the \mbox{HFS-resolved} rate coefficients for inelastic collisions of HCN  with \mbox{para-H$_2$}
and  \mbox{ortho-H$_2$} (the later dominate when \mbox{$T_{\rm k}$\,$>$\,40\,K}
if the H$_2$ OPR is thermalized). We computed these coefficients using the \mbox{S-IOS} approximation up to \mbox{$J$\,=\,11} 
 and  \mbox{$T_{\rm k}$\,=\,500\,K}. We also studied the
role of \mbox{HCN--e$^-$} collisions using  \mbox{HFS-resolved}  rates 
of \cite{Faure07}.

We ran a grid of static and uniform  cloud models  
appropriate to the physical conditions  in GMCs. As found by previous studies, line overlap and opacity effects alter the HFS level populations and the emitted HCN rotational  spectrum when lines become optically thick, roughly at \mbox{$N$(HCN)\,$>$\,10$^{13}$\,cm$^{-2}$}.
As a result, the  relative \mbox{$J$\,=\,1--0} HFS line intensity  ratios ($R_{12}$ and $R_{02}$) deviate from the \mbox{optically} thin limit ratios \mbox{1:5:3} (\mbox{$R_{12}$\,=\,0.6} and \mbox{$R_{02}$\,=\,0.2}). \mbox{Anomalous} intensity ratios imply different excitation temperatures and often different line widths for each HFS line 
 (the basic assumption of automatic HFS line-fitting programs). In addition, the  \mbox{$J$\,=\,2--1} and \mbox{3--2} HFS spectra can be shown to be anomalous as well. 
 Our models reproduce the anomalous HCN \mbox{$J$\,=\,1--0}  spectra observed in the \mbox{Orion Bar} and Horsehead PDRs.

 As shown in previous studies focused on the HCN \mbox{rotational} excitation alone,  electron collisions become important for H$_2$ densities below a few 
10$^5$\,cm$^{-3}$ and electron abundances \mbox{$\chi_{\rm e}$\,$>$\,10$^{-5}$}. 
  Electrons and line overlap effects enhance the excitation of  
higher $J$ rotational levels. They enhance the emitted rotational line intensities even if the hyperfine structure is not resolved.   
We show that also when electron collisions dominate,
the HCN \mbox{$J$\,=\,1--0} HFS spectrum can be anomalous.
In these cases, electron excitation increases the \mbox{$J$\,=\,1--0} HFS line intensities by up to
an order of magnitude if $\chi_{\rm e}$\,$\simeq$\,10$^{-4}$, and can
 produce low-surface-brightness HCN emission from the
low-density-gas \mbox{(several \mbox{10$^3$\,cm$^{-3}$})} extended environment of GMCs
(tens of pc).  
The \mbox{ubiquity} of such an extended HCN emission component in GMCs, if confirmed, may affect the 
interpretation of the spatially unresolved extragalactic HCN emission, which may not always
be dominated by dense (\mbox{$>$\,10$^5$\,cm$^{-3}$}) molecular gas \mbox{-- at} least in normal galaxies in which $L_{\rm HCN}$ correlates
with  $L_{\rm CO}$.
\mbox{Alternatively}, the extended HCN emission in GMCs might
be line photons emitted by dense molecular cores and resonantly scattered over wide
spatial scales by large envelopes of lower density gas. Both scenarios produce
different $R_{12}$ and $R_{02}$ ratios, which are more anomalous in the scattering 
envelope case, but currently less frequently seen in  GMCs observations.
Thus, the two scenarios should be tested based on observations of large-scale HCN HFS emission in galactic GMCs.
All in all, we expect that a  proper excitation analysis of ongoing HCN  emission surveys   will constrain the
dominant origin and physical conditions of the HCN emitting gas in GMCs,
as well as its relation with the extragalactic star formation rate correlations.

\begin{acknowledgements}  
We thank A. Faure for sharing his \mbox{HCN-HFS + $e$}  rate coefficients in tabulated form. We warmly thank 
\mbox{S. Cuadrado}, \mbox{J. Pety}, and \mbox{M. Gerin}   for providing the HCN  \mbox{$J$\,=\,1--0} spectra of the Orion Bar and Horsehead,
 and for  useful discussion on the HCN \mbox{$J$\,=\,1--0} HFS emission
in Orion~B.  We thank our referee for concise and illuminating comments.
 JRG and MGSM thank the Spanish MCINN for funding support under grant
\mbox{PID2019-106110GB-I00}.

\end{acknowledgements}

%
%

\bibliographystyle{aa}
\bibliography{references}

\begin{appendix}\label{Sect:Appendix}

\section{HCN HFS energy diagram}

\begin{figure}[t]
\centering   
\includegraphics[scale=0.38, angle=0]{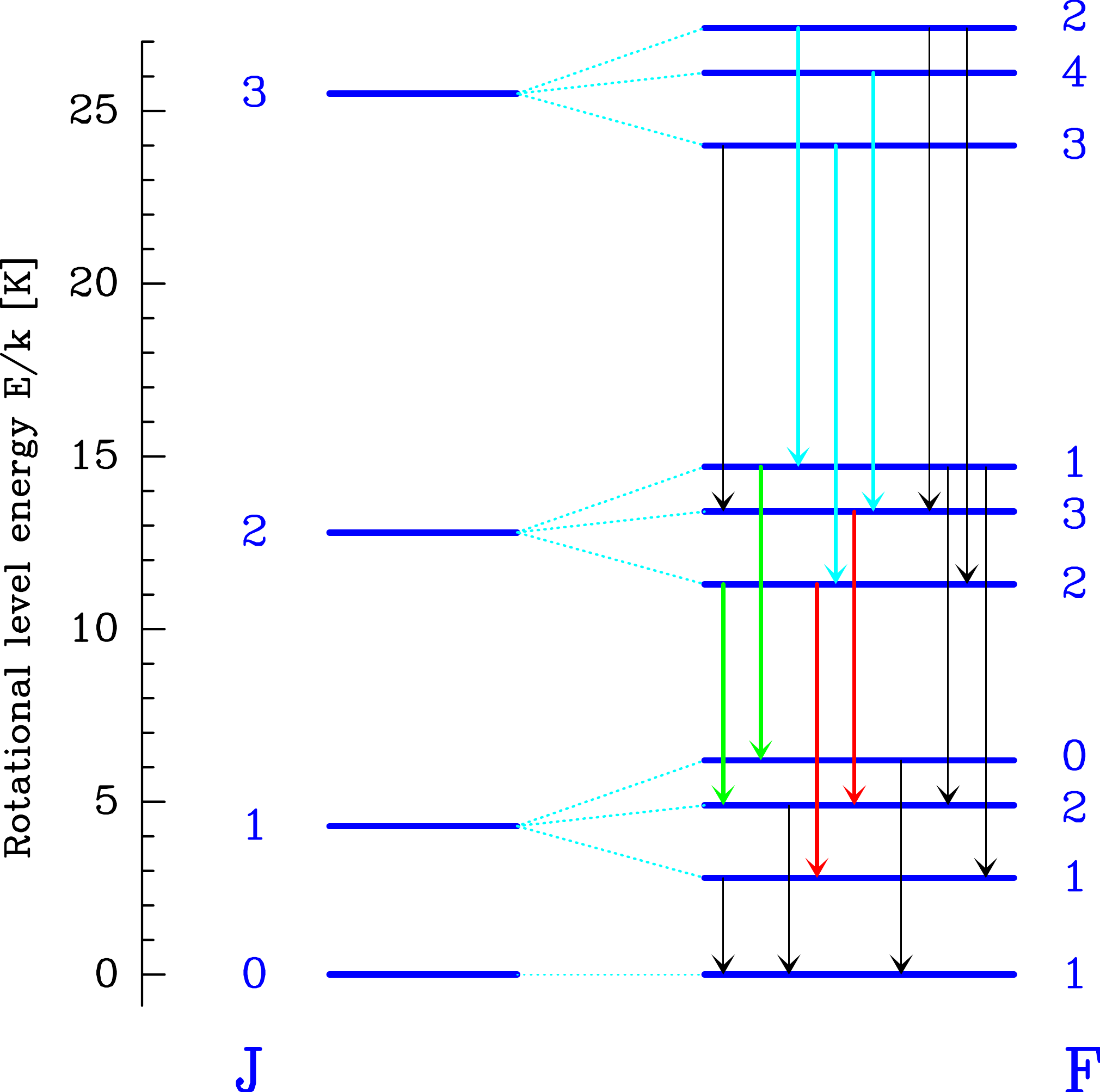}
\caption{HCN  rotational energy diagram and HFS splittings (\mbox{exaggerated} for clarity).
Arrows show dipole-allowed radiative transitions. Red, green, and cyan  arrows show
lines that overlap, in interstellar conditions, and produce 
most of the \mbox{anomalous} HFS emission discussed in the text.} 
\label{fig:energy_diagram}
\end{figure}

Figure~\ref{fig:energy_diagram} shows the low-lying HCN rotational levels and HFS structure taken
from CDMS \citep{Endres16} using spectroscopic data from \cite{Ahrens02,Thorwirth03}, and references therein.

\section{Absolute line intensities}\label{App-intensities}

\begin{figure*}[t]
\centering   
\includegraphics[scale=0.69, angle=0]{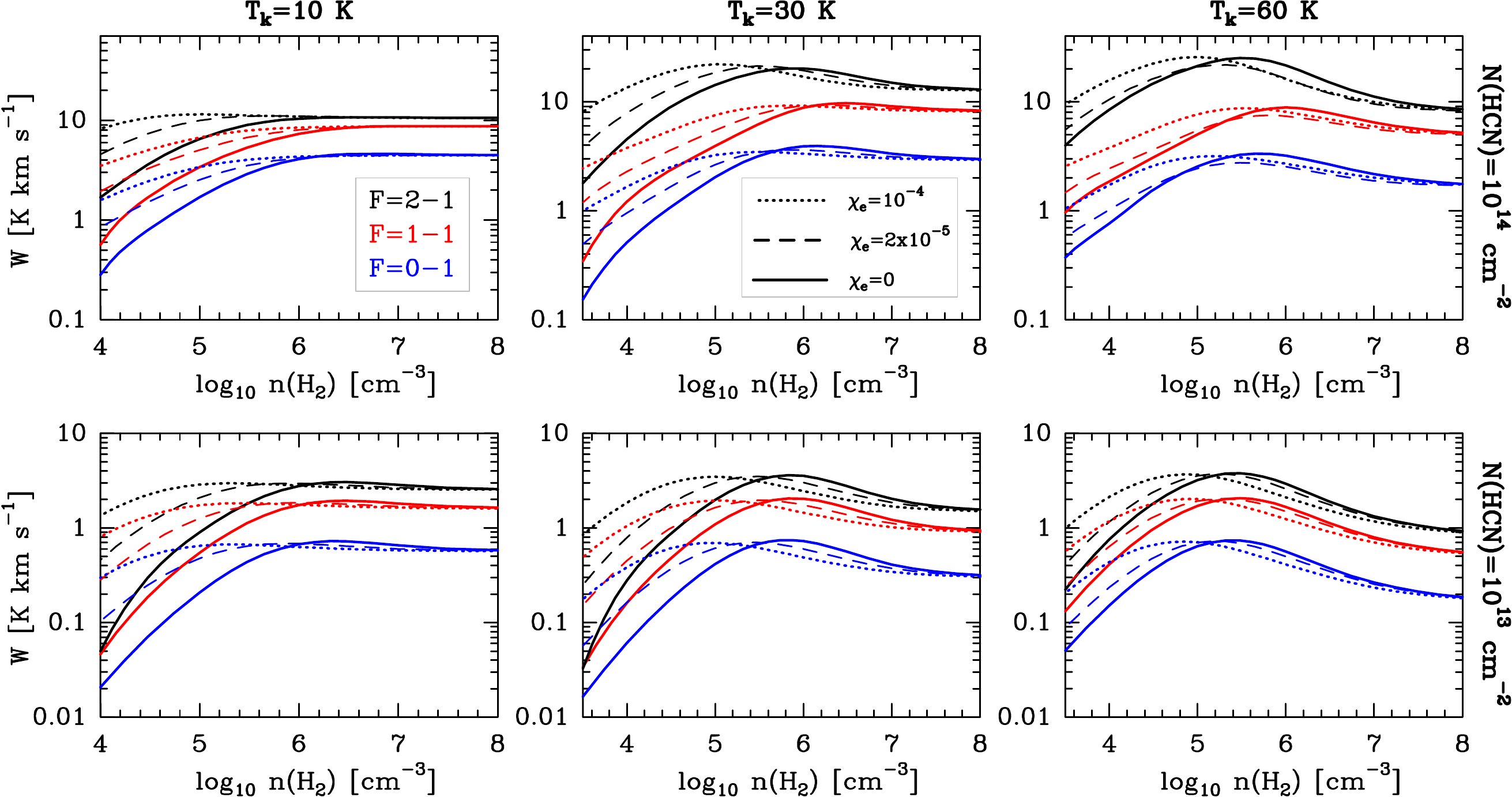}
\caption{\mbox{HCN $J$\,=\,1--0}  HFS integrated line intensities; 
$I$($F_{\rm u}$--$F_{\rm l}$) in K\,km\,s$^{-1}$.
Model results show the role of electron  excitation for different ionization fractions: 
\mbox{$\chi_{\rm e}$\,=\,0} (continuous curves), $\chi_{\rm e}$\,=\,2$\cdot$10$^{-5}$ (dashed), and 
$\chi_{\rm e}$\,=\,10$^{-4}$ (dotted). All models include line overlap.} 
\label{fig:grid_W_xe}
\end{figure*}

Figure~\ref{fig:grid_W_xe} shows absolute intensities of the HCN~$J$\,=\,1--0, 2--1, and 3--2 lines
(integrating over all HFS components) from our grid of static cloud models including line overlap effects
and three different electron abundances: \mbox{$\chi_{\rm e}$\,=\,0} (continuous curves), 
\mbox{$\chi_{\rm e}$\,=\,2$\cdot$10$^{-5}$} (dashed), and  \mbox{$\chi_{\rm e}$\,=\,10$^{-4}$} (dotted).

\section{Recoupling vs. S-IOS  HCN--H$_2$ HFS rate coefficients}\label{App-recoupling-IOS}

\begin{figure}[t]
\centering   
\includegraphics[scale=0.75, angle=0]{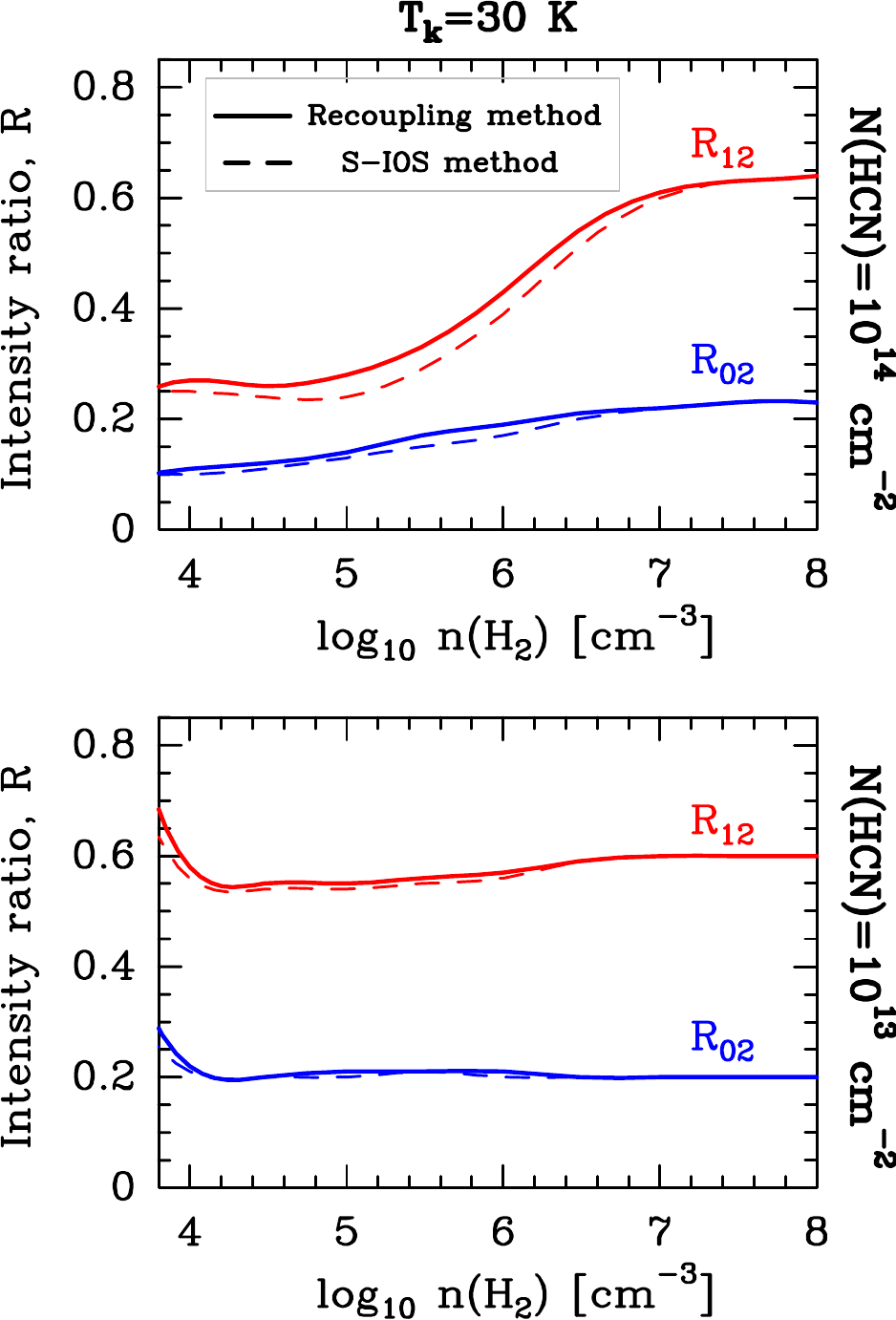}
\caption{ \mbox{$J$\,=\,1--0}  HFS integrated line intensity ratios \mbox{$R_{12}$\,=\,$I$($F$=1--1)\,/\,$I$($F$=2--1)}
and \mbox{$R_{02}$\,=\,$I$($F$=0--1)\,/\,$I$($F$=2--1)} using different methodologies to compute the collisional rate coefficients at
\mbox{$T_{\rm k}$\,=\,30\,K}: recoupling (continuous curves) and S-IOS (dashed lines).}
\label{fig:grid_tk30_IOS_CC}
\end{figure}

Figure~\ref{fig:grid_tk30_IOS_CC} shows the effects of using the 
\mbox{H$_2$--HCN} HFS collisional rate coefficients computed in the \mbox{S-IOS}
approximation (dashed lines) and using rates in the nearly exact recoupling method (continuous curves)
of \cite{Magalhes:18}.
These are cloud  models at \mbox{$T_{\rm k}$\,$=$\,30\,K} (H$_2$ \mbox{OPR\,$\simeq$\,0.03}).
Thus, they only consider collisions with \mbox{para-H$_2$}.
 For optically thin or slightly
thick lines (models with $N$(HCN)\,=\,10$^{13}$\,cm$^{-2}$) line overlap effects are very minor, the resulting $R_{12}$ and $R_{02}$ ratios using different collisional rate datasets are nearly identical (bottom panel in Fig.~\ref{fig:grid_tk30_IOS_CC}).
For higher $N$(HCN) and $\tau_{\rm hfs}$, line overlap effects become important. Despite the slightly different
rate coefficients, especially the quasi-elastic rates ($\Delta$$J$\,=\,0), the $R_{12}$ and $R_{02}$ ratios
are only (up to) $\sim$\,15\,$\%$ lower than in models that use the more approximated  \mbox{S-IOS} rates.
The absolute line intensities only differ by a few percent.
Therefore, we conclude that the adoption of rates  calculated in the \mbox{S-IOS} approximation at high $T_{\rm k}$ should be
accurate enough in most astrophysical applications. Hence, it benefits from the use of HFS-resolved  rates for  collisions with both \mbox{ortho-H$_2$} and \mbox{para-H$_2$} and at higher gas temperatures.

\end{appendix}

\end{document}